\documentclass[%
  reprint,
 amsmath,amssymb,longbibliography
 aps,
prd]{revtex4-2}

\usepackage{graphicx}
\usepackage{dcolumn}
\usepackage{bm}
\usepackage{lipsum}
\usepackage{appendix}
\usepackage{siunitx}
\usepackage{glossaries}
\usepackage{todonotes}
\usepackage{dirtytalk} % quotes
\usepackage{multirow}
\usepackage{booktabs}
\usepackage{rotating}

\usepackage{hyperref}
\hypersetup{
    colorlinks=true,
    linkcolor=red,
    anchorcolor=black,
    citecolor=blue,
    filecolor=cyan,
    menucolor=red,
    runcolor=cyan,
    urlcolor=magenta,
    pdftitle={Stimulated Brillouin scattering by surface acoustic waves in lithium niobate waveguides},
    pdfpagemode=FullScreen,
    }
\usepackage{cleveref}
\Crefname{equation}{Eq.}{Eqs.}
\Crefname{figure}{Fig.}{Figs.}
\Crefname{table}{Tab.}{Tabs.}

\begin{document}

\title{Stimulated Brillouin scattering by surface acoustic \\ waves in lithium niobate waveguides}

\author{Caique C. Rodrigues}
\affiliation{\vspace{0.25cm} Gleb Wataghin Physics Institute, University of Campinas, Campinas, SP, Brazil \\
Photonics Research Center, University of Campinas, Campinas, SP, Brazil}%

\author{Roberto O. Zurita}
\affiliation{\vspace{0.25cm} Gleb Wataghin Physics Institute, University of Campinas, Campinas, SP, Brazil \\
Photonics Research Center, University of Campinas, Campinas, SP, Brazil}%

\author{Thiago P. M. Alegre}
\affiliation{\vspace{0.25cm} Gleb Wataghin Physics Institute, University of Campinas, Campinas, SP, Brazil \\
Photonics Research Center, University of Campinas, Campinas, SP, Brazil}%

\author{Gustavo S. Wiederhecker}
\email{gsw@unicamp.br}
\affiliation{\vspace{0.25cm} Gleb Wataghin Physics Institute, University of Campinas, Campinas, SP, Brazil \\
Photonics Research Center, University of Campinas, Campinas, SP, Brazil}%

\begin{abstract}
    \textbf{Abstract:} We numerically demonstrate that Lithium Niobate on Insulator (LNOI) waveguides may support confined short-wavelength surface acoustic waves that interact strongly with optical fields through backward stimulated Brillouin scattering in both $Z$ and $X$-cut orientation. We conduct fully anisotropic simulations that consider not only moving boundary and photoelastic forces, but also roto-optic forces for the Brillouin interaction. Our results indicate that photoelasticity dominates the Brillouin gain and can reach as high as $G_{B}/Q_{m}$ = 0.43 W$^{-1}$m$^{-1}$ in standard ridge waveguides.
\end{abstract}

\maketitle

\section{Introduction}

Stimulated Brillouin scattering (SBS) was first observed in 1964~\cite{PhysRevLett.12.592} in quartz and sapphire and later in liquids~\cite{PhysRevLett.13.334}. The exploration of SBS in piezoelectric materials and other lower symmetry crystals was tackled soon after, with early observations~\cite{doi:10.1063/1.1655749} followed by theoretical treatment~\cite{PhysRevB.31.1034}. The development of optical fibers led to a wide range of SBS applications, from distributed sensing to fiber lasers~\cite{doi:10.1063/1.1654249, doi:10.1063/1.88583, doi:10.1063/1.90726,doi:10.1063/1.89965}. In recent times, SBS has been observed in strongly confining chalcogenide glass (As$_{2}$S$_{3}$)~\cite{Pant:11}, silicon~\cite{Shin2013, VanLaer2015} and silicon nitride waveguides~\cite{PhysRevLett.124.013902,doi:10.1126/sciadv.abq2196}. Lithium niobate (LiNbO$_{3}$, or LN), despite its long history in bulk and weakly confining optical waveguides and acousto-optic devices, has only recently been studied for high-confinement integrated waveguides and cavities, enabling enhanced interaction between optical and mechanical waves~\cite{Jiang2016,Jiang:19,https://doi.org/10.48550/arxiv.2210.01064}. The challenge of tailoring LN structures to enhance optomechanical interaction lies in achieving simultaneous confinement of optical and mechanical waves. Existing strategies either rely on fully suspended structures or the exploration of surface acoustic waves in LN. However, in these demonstrations the acoustic waves are excited through surface-attached electric transducers that exploit the piezoelectric properties of LN. 

In this study, we present a Lithium Niobate on Insulator (LNOI) integrated waveguide that guides light and short-wavelength surface acoustic waves simultaneously, without relying on suspended structures. Full optical and mechanical anisotropy are included to boost the SBS gain. Our investigation begins by examining how LN's anisotropy enables the engineering of its photoelasticity using finite element method (FEM). We then simulate an idealized structure, a floating waveguide, as a benchmark for future simulation. Subsequently, we examine two common waveguide geometries, the strip and the ridge waveguides~\cite{PhysRevA.92.013836,Eggleton2019,Poulton:13,Eggleton:13,PhysRevA.93.053828,JESipe_2016,Wolff:21,doi:10.1063/1.5088169}.

\section{Overview of SBS in LN\label{sec:eng_photo}}

\begin{figure}[h!]
    \centering
    \includegraphics[width=6cm]{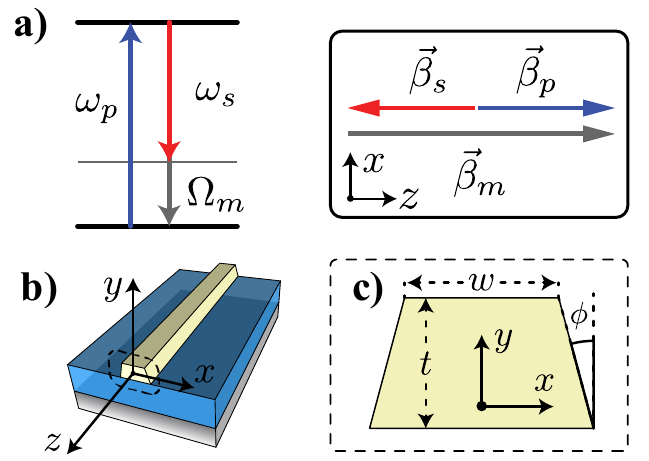}
    \caption{\textbf{a)} Backward Brillouin scattering process, illustrating the conservation of energy $\Omega_{m} = \omega_{p} - \omega_{s}$ and the phase-matching $\vec{\beta}_{m} = \vec{\beta}_{p} - \vec{\beta}_{s} \approx 2\vec{\beta}_{p}$; \textbf{b)} 3D illustration of the system of interest, which is a LN waveguide on silicon dioxide (silica, SiO$_{2}$). The axes are the laboratory coordinates used, which is always written in lower case letters; \textbf{c)} Geometrical parameters of the waveguide's cross-section.}
    \label{fig:LN_SBS_Waveguide}
\end{figure}

In stimulated backward Brillouin scattering, an optical pump wave propagating through a waveguide with angular frequency $\omega_{p}$ and wavevector $\vec{\beta}_{p}$ is backscattered by thermally excited acoustic phonons, creating frequency-shifted backscattered waves with higher (anti-Stokes) and lower (Stokes) frequency components, shifted precisely by the acoustic phonon frequency ($\Omega_{m}$). The interference between the pump and the backscattered Stokes light (which has a wavevector $\vec{\beta}_{s} \approx -\vec{\beta}_{p}$ and angular frequency $\omega_{s}$) has a beat note at $\Omega_{m} = \omega_{p} - \omega_{s}$. This optical beating acts as a source for acoustic waves obeying the phase-matching condition $\vec{\beta}_{m} = \vec{\beta}_{p} - \vec{\beta}_{s} \approx 2\vec{\beta}_{p}$, where $\vec{\beta}_{m}$ is the acoustic wavevector. This interaction and the target LNOI waveguide geometry are illustrated in \Cref{fig:LN_SBS_Waveguide}.

The strength of the Brillouin interaction in waveguides is characterized by the normalized Brillouin gain $G_{B}/Q_{m}$, in which $G_{B}$ is the Brillouin gain [W$^{-1}$m$^{-1}$] and $Q_{m}$ is the total mechanical quality factor. The peak Brillouin gain can be calculated from the spatial overlap between optical and acoustic modes~\cite{doi:10.1063/1.5088169},

\begin{equation}{\label{eq:gain}}
    \frac{G_{B}}{Q_{m}} = \frac{2\omega_{p}}{m_{\text{eff}}\Omega_{m}^{2}}\left|\int \left(f_{\text{PE}}+f_{\text{RO}}\right)dA + \int f_{\text{MB}}dl\right|^{2},
\end{equation}

\noindent where $m_\text{eff}$ is the effective linear mass density, $f_\text{PE},f_\text{RO}$ and $f_\text{MB}$ are the photoelastic, roto-optic, and moving boundary contributions to the optical stress densities (expressions are defined in \Cref{optical_force_densities}). The magnitude of such stress densities depends not only on the spatial distribution of optical and acoustic modes, but also on the magnitude of intrinsic material parameters. Namely, $f_\text{PE}$ increases with refractive index and photoelastic parameters; $f_\text{RO}$ depends on the degree of optical anisotropy, while $f_\text{MB}$ depends on the dielectric permittivity contrast between interfaces. 

Despite the complexity of \Cref{eq:gain}, an approximate expression for the backward SBS gain can be derived under some assumptions. For instance, in a weakly guiding optical waveguide interacting with a plane-wave like longitudinal acoustic mode, the Brillouin gain is dominated by photoelasticity and can be simplified to,

\begin{equation}\label{eq:optimal_GB}
    \frac{G_{B}}{Q_{m}} \approx \frac{2\omega_{p}}{m_{\text{eff}}\Omega_{m}^{2}}\frac{n_{\text{eff}}^{6}p_{\text{eff}}^{2}\beta_{m}^{2}}{4c^{2}} = \frac{\pi}{c}\frac{n^{6}_{\text{eff}}p^{2}_{\text{eff}}}{m_{\text{eff}}V_{L}^{2}}\frac{1}{\lambda_{p}},
\end{equation}

\noindent where $n_\text{eff}$ is effective refractive index, $p_{\text{eff}}$ is an effective photoelastic coefficient, $c$ is the speed of light, $\lambda_{p} = 2\pi c/\omega_{p}$ is the pump wavelength, $m_{\text{eff}} = \rho A_{\text{eff}}$ where $\rho$ is the material density and $A_\text{eff}$ is the effective optical mode area. The mechanical wavenumber $\beta_{m} = \Omega_{m}/V_{L}$ in \Cref{eq:optimal_GB}, where $V_{L}$ is the acoustic longitudinal velocity, follows from the assumption that the dominant strain component lies along the propagation direction for a longitudinal acoustic mode, thus, the accompanying photoelastic parameter $p_{\text{eff}}$ is often taken as $p_{13}$, which for an isotropic material (e.g., silica) is equal to $p_{12}$. As shown in \Cref{fig:LN_Slowness_and_PIJ}, LN is a highly anisotropic material, if the orientation of the crystal is changed with regards to the propagation direction, the photoelastic coefficients may vary substantially depending on the direction of the waveguide along the laboratory $(x,y)$-plane. If we naively use typical LN waveguide parameters (shown in \Cref{tab:LN_typical_params}) and assume either $p_{11} \approx -0.026$ or $p_{12} \approx 0.134$ as the dominant photoelastic component (for the photoelastic tensor written in the crystal's principal axes), the normalized Brillouin gain predicted by \Cref{eq:optimal_GB} will vary more than an order of magnitude between $0.006<G_{B}/Q_{m}<0.16$ in units of W$^{-1}$m$^{-1}$. The wide range covered by this simple estimate suggests that LN anisotropy cannot be ignored when exploring the SBS interaction. Indeed, the upper bound suggests that there could be specific crystal orientations where LN waveguides could experience large Brillouin gain, even when compared to silicon~\cite{VanLaer2015} and chalcogenide~\cite{Pant:11} integrated waveguides. Meanwhile, previous simulations articles did not properly consider LN anisotropy~\cite{app11188390}. Previous cavity optomechanics experiments with LN structures~\cite{Jiang:19} indicate the importance of properly handling its anisotropic properties, yet, SBS in LN waveguides has not yet been experimentally observed. The numerical calculations of this work were all done using \Cref{eq:gain}, while the discussion using \Cref{eq:optimal_GB} is more of a motivation, as well a figure of merit analysis.

Thus, we present a comprehensive, full-vectorial numerical study of SBS in LN waveguides that addresses two crucial questions. On one hand, we demonstrate that surface acoustic waves (SAWs) in both $Z$-cut and $X$-cut LN can attain high normalized Brillouin gains ($G_{B}/Q_{m}$) if the proper crystal orientations are chosen. On the other hand, utilizing SAWs to attain large gains eliminates the need for suspended waveguides as they remain guided even in the presence of a silicon dioxide substrate. This is remarkable as the buried oxide has a shear acoustic velocity slightly lower than that of LN, but a higher longitudinal acoustic velocity, preventing total internal reflection for hybrid bulk acoustic waves (BAWs), as depicted in \Cref{fig:LN_Slowness_and_PIJ}(b,e).

\section{Anisotropy of LN\label{sec:anisotropy}}

\begin{figure*}[ht]
    \centering
    \includegraphics[width=16cm]{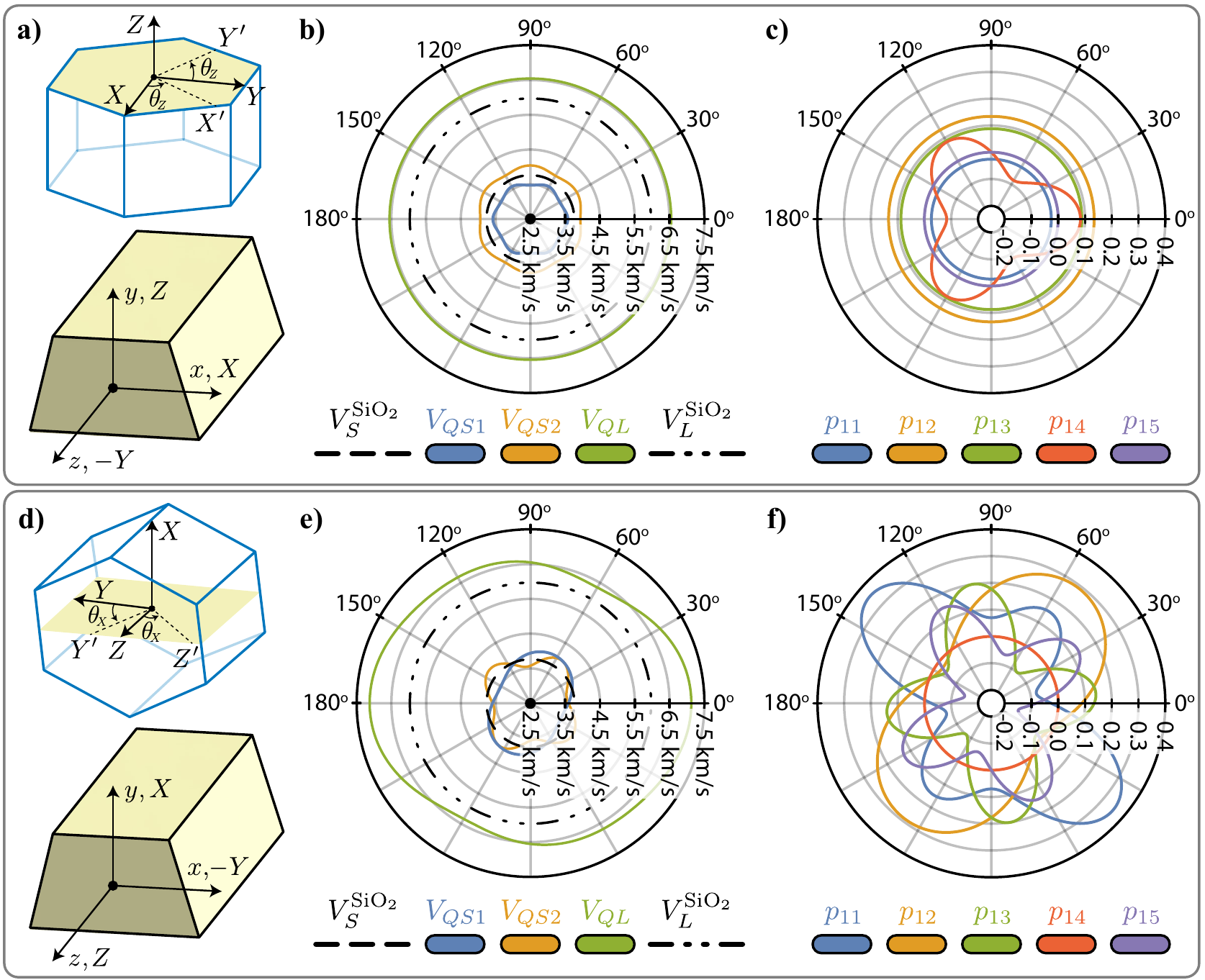}
    \caption{\textbf{a)}-\textbf{d)} Crystallographic orientation ($X,Y,Z$) relative to the laboratory coordinate axes ($x,y,z$) for the $Z$-cut and $X$-cut, respectively, as well the definition of the rotation angles $\theta_{Z}$ and $\theta_{X}$; \textbf{b)}-\textbf{e)} Bulk LN velocities curves in both $Z$ and $X$ cuts, respectively. $V_{QS1}$ and $V_{QS2}$ - Quasishear LN velocities 1 and 2, respectively; $V_{QL}$ - Quasilongitudinal LN velocity. The shear ($V_{S}$) and longitudinal ($V_{L}$) velocities for silica are also shown in dashed circles for comparison; \textbf{c)}-\textbf{f)} Polar plot of some of the photoelastic coefficients $p_{IJ}$ written in the laboratory axes as we rotate $\theta_{Z}$, in the $Z$-cut, and $\theta_{X}$, in the $X$-cut, respectively.}
    \label{fig:LN_Slowness_and_PIJ}
\end{figure*}

The large optical and mechanical anisotropy of LN cannot be ignored in the SBS interaction. LN is a trigonal crystal where niobium oxide and lithium are stacked in hexagonal plane sheets along the $c$-axis (or $Z$-axis, following the convention that capital letters represent the material axes)~\cite{Weis1985,ABRAHAMS1966997, doi:10.1063/1.1660528,Yamada_1967,Jazbinsek2002, Zhu:21,LEDBETTER2004941,doi:10.1121/1.385588}. \Cref{fig:LN_Slowness_and_PIJ} reveals the anisotropic characteristic of the material and clarifies the distinction between laboratory ($x,y ,z$) and material coordinate ($X,Y,Z$) systems, which is important to define the appropriate tensorial properties transformations. The SBS interactions in an anisotropic material will depend on the permittivity tensor $\varepsilon_{ij}$, the lossless stiffness tensor $C_{ijkl}$, and the photoelastic tensor $p_{ijkl}$. In our modal analysis we will ignore second order effects arising from the electro-optic, piezoelectric, and Pockels nonlinearity, although they could play a role in piezo-induced mechanical losses and higher order nonlinearities. 

The two most common orientations for LNOI wafers are the $X$-cut and $Z$-cut, as illustrated in \Cref{fig:LN_Slowness_and_PIJ}(a,d). In the more symmetric $Z$-cut, the waveguide top surface normal ($\hat{y}$) is aligned with the crystal's $c$-axis ($\hat{Z}$). This orientation exhibits a three-fold rotational ($\theta_{Z}$) symmetry around the $Z$ (crystal) or $y$ (laboratory) axis. In contrast, the $X$-cut wafer has the $c$-axis orthogonal to the waveguide top surface normal and exhibits only a two-fold rotational symmetry. It is worth noting that many of the relevant material properties exhibit full rotational symmetry for $Z$-cut, as shown in \Cref{fig:LN_Slowness_and_PIJ}(b,c), while the $X$-cut has more substantial variations of these parameters as the structure rotates around the $X$ (crystal) or $y$ (laboratory) axis, as shown in \Cref{fig:LN_Slowness_and_PIJ}(e,f). The photoelastic coefficients $p_{12},p_{13},$ and $p_{14}$, shown in \Cref{fig:LN_Slowness_and_PIJ}(c,f), weight the dominant strain components present in $yz$-polarized surface acoustic waves, revealing a strong dependence on orientation. It is also worth comparing the LN acoustic velocities with silicon dioxide, the target substrate. LN’s longitudinal velocity is always higher, preventing pressure-like waves commonly explored in SBS from being guided in the LN layer. However, for surface waves, which propagate at lower speeds, \Cref{fig:LN_Slowness_and_PIJ}(b,e) indicates that there could be acoustic guidance.

\section{Numerical Analysis\label{sec:numerical}}

To investigate the properties predicted from the earlier analysis, we examine the typical geometry of high confinement LN waveguides shown in \Cref{fig:LN_SBS_Waveguide}(b,c). The waveguide's width and thickness chosen in this study fall within the range of previous experiments on LN waveguides~\cite{Desiatov:19,Wang:17,https://doi.org/10.48550/arxiv.2208.05556}. We employ the finite element method (FEM) using COMSOL Multiphysics to calculate optical and mechanical modes of our structures. All calculations consider all factors in \Cref{eq:gain} (i.e., $f_{\text{PE}}$, $f_{\text{RO}}$ and $f_{\text{MB}}$), as well as the full anisotropy of LN. The laboratory coordinate system is fixed and oriented as shown in \Cref{fig:LN_Slowness_and_PIJ}(a) or \Cref{fig:LN_Slowness_and_PIJ}(d), with  direction of optical and acoustic wave propagation along the $z$-axis. To study the crystal orientation dependence, either a $X$ or $Z$-cut is selected, and the geometries explored are rotated around the laboratory $y$-axis.

\subsection{SBS in $Z$-cut LN\label{sec:SBS_zcut}}

\begin{figure*}[ht]
    \centering
    \includegraphics[width=16cm]{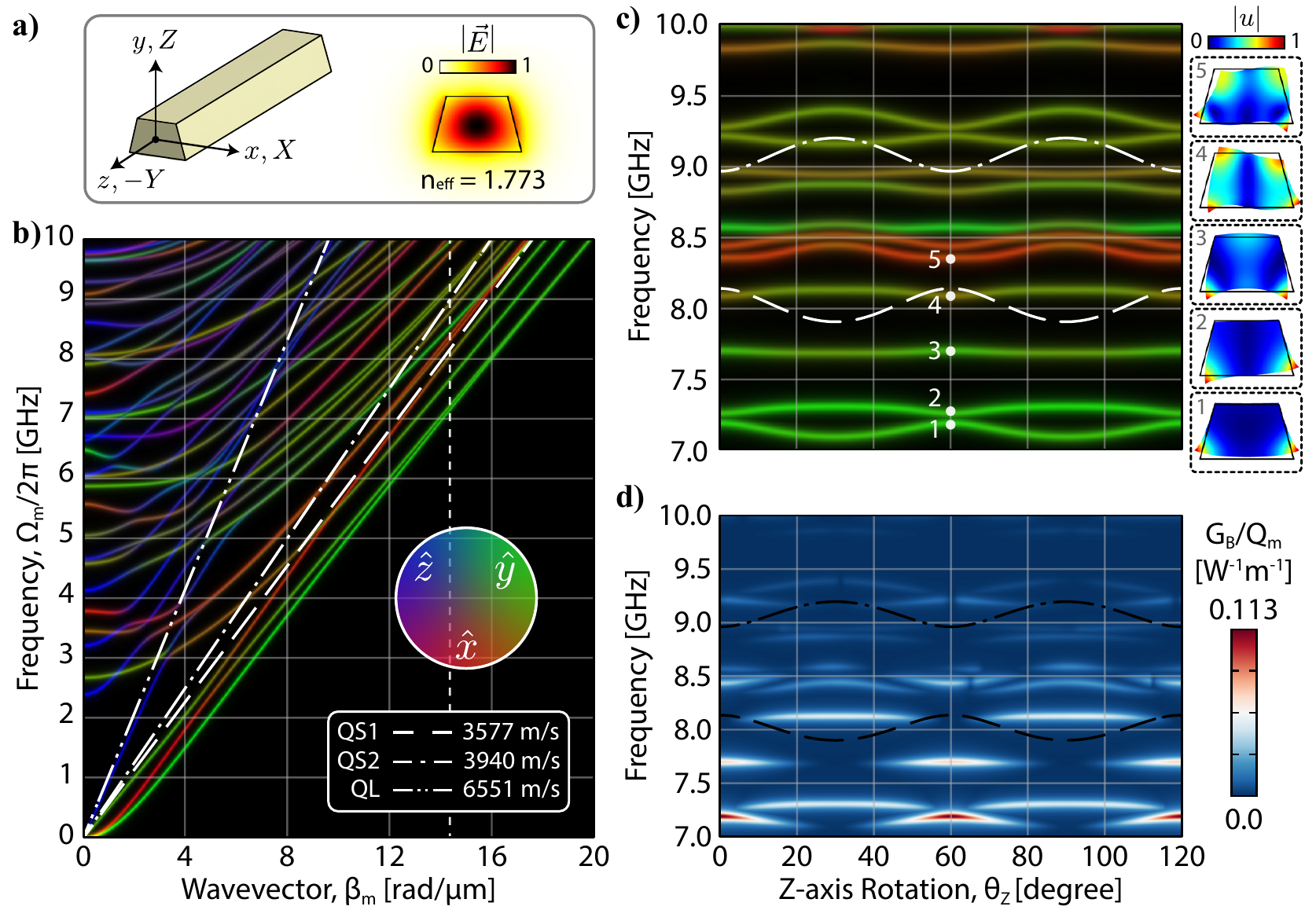}
    \caption{\textbf{a)} Initial relative orientation between material ($X,Y,Z$) and coordinate ($x,y,z$) axes, as well the optical mode profile with normalized electric field $|\vec{E}|$ used to calculate the Brillouin gain; \textbf{b)} $Z$-cut mechanical dispersion analysis. The colors represent a weighting of mechanical polarization, where we have chosen pure blue for longitudinal waves ($z$-polarized waves), pure green for $y$-polarized waves and pure red for $x$-polarized waves, as defined in \Cref{sec:color_mechpol}. We used $\theta_{Z}=0^{\circ}$ for this data; \textbf{c)} $Z$-cut mechanical dispersion analysis as a function of $\theta_{Z}$ with fixed $\beta_{m} = 14.37$ $\mu$m$^{-1}$. The dashed curves are the solutions of the Christoffel equation for an infinite LN medium. The first five mechanical modes profiles with normalized acoustic displacement field $|\vec{u}|$ are shown; \textbf{d)} $Z$-cut normalized Brillouin gain density plot as a function of $\theta_{Z}$. All data were simulated using the floating waveguide with $w=650$ nm, $t=550$ nm and $\phi=15^{\circ}$.}
    \label{fig:LN_Z-cut_Dispersions_and_GB}
\end{figure*}

We begin by analyzing the optical and mechanical dispersion independently for a given crystallographic $Z$-cut orientation in a floating waveguide, as illustrated in \Cref{fig:LN_Z-cut_Dispersions_and_GB}(a). Unlike the anchored LNOI waveguide shown in \Cref{fig:LN_SBS_Waveguide}(b), this suspended geometry allows us to obtain the mechanical dispersion relation and understand the presence of surface waves without the influence of the silicon dioxide substrate. Moreover, the suspended structure provides the maximum level of optical and mechanical confinement, making it a reference for the Brillouin gain in a typical LN waveguide.

For the $Z$-cut orientation, the optical permittivity has full rotation symmetry around the $y$-axis. Thus, we fix the optical mode at the wavelength $\lambda_{p} = 1550$ nm, as shown in \Cref{fig:LN_Z-cut_Dispersions_and_GB}(a). The effective index of $n_\text{eff}=1.773$ leads to a backward SBS phase-matched mechanical wavevector $\beta_{m}=4\pi n_\text{eff}/\lambda_{p}=14.37$ $\mu$m$^{-1}$. The mechanical dispersion shown in \Cref{fig:LN_Z-cut_Dispersions_and_GB}(b) reveals a high density of modes with different mechanical polarizations, as visualized by our color identification. The degree of polarization for each spatial direction is calculated as in~\cite{Zurita:21}. The bulk acoustic velocities for this orientation of LN are also shown in the figure; with the slowest mechanical modes (below the QS1 line) exhibiting SAW speed characteristics. Their slow phase velocities could ensure guidance in the presence of a silicon dioxide substrate. We calculated the mechanical frequencies for these modes (fixed $\beta_{m}$) as a function of the rotation angle around the $Z$-axis, $\theta_{Z}$, and the results, along with the spatial displacement profile for the five lowest frequency modes, are shown in \Cref{fig:LN_Z-cut_Dispersions_and_GB}(c). Modes 1 and 2 are localized in the lower corners of the suspended LN structure, so they are not of interest to us, as they should be suppressed in the presence of a substrate. However, mode 3 shows promising displacement in the central waveguide region. \Cref{fig:LN_Z-cut_Dispersions_and_GB}(d) displays the full-vectorial SBS gains ($G_{B}$) normalized by the mechanical quality factor ($Q_{m}$), revealing that the lowest frequency modes may have reasonable normalized SBS gains, provided the appropriate direction is chosen. The maximum normalized gain of $G_{B}/Q_{m}=0.113$ W$^{-1}$m$^{-1}$ is about 30\% of the normalized gain observed for backward SBS in suspended silicon waveguides ($G_{B}/Q_{m} \approx 0.37$ W$^{-1}$m$^{-1}$~\cite{VanLaer2015}). The maximum SBS gains are obtained along the same orientation that results in the maximum negative value for $p_{14} \approx -0.1$. This gain is dominated by photoelastic strain induced by $yz$ shear strain component, as expected for the vertical shear polarization of the modal profile, and this is verified by the relative contributions from each effect in \Cref{supp} \Cref{fig:LN_PE_MB_RO}(a,b,c).

\subsubsection{Strip Waveguide $Z$-cut LN\label{sec:zstrip}}

\begin{figure}[hb]
    \centering
    \includegraphics[width=7cm]{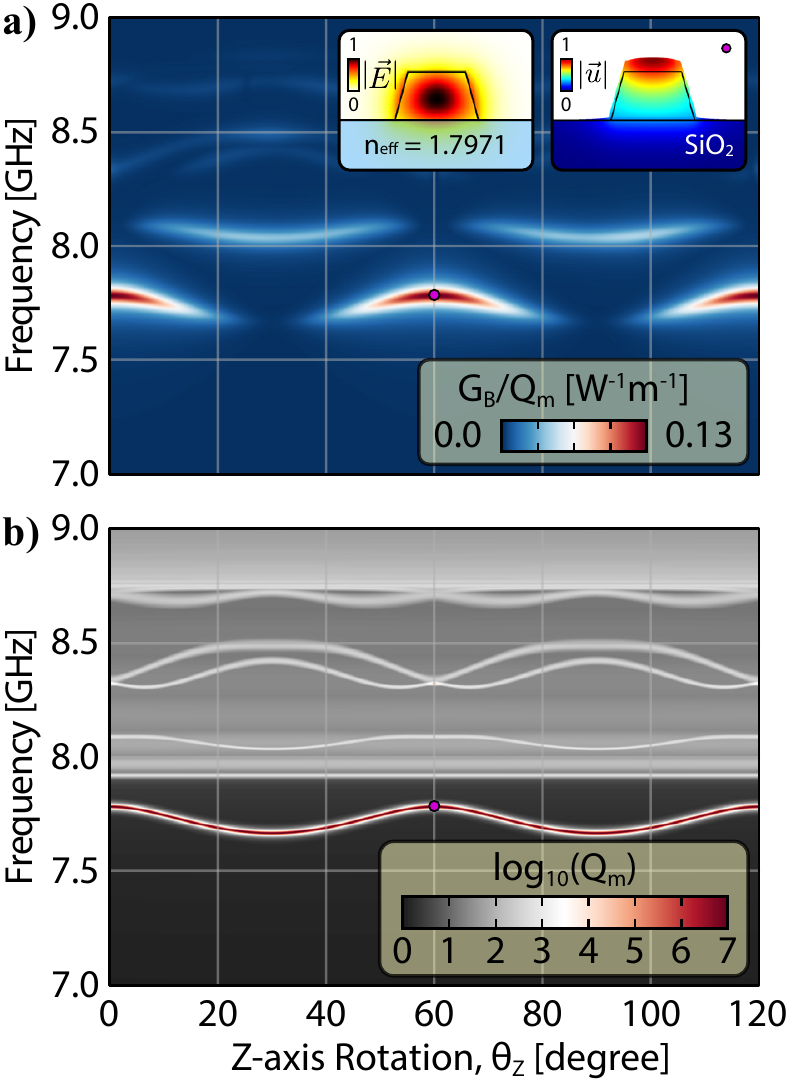}
    \caption{\textbf{a)} $Z$-cut normalized Brillouin gain $G_{B}/Q_{m}$ density plot as a function of $\theta_{Z}$; \textbf{b)} Mechanical quality factor $Q_{m}$ density plot as a function of $\theta_{Z}$. We truncated log$_{10}\left(Q_{m}\right)$ at 7 to not saturate the map. The dot in purple at $\theta_{Z}=60^{\circ}$ corresponds to the fundamental mechanical mode at the inset.}
    \label{fig:LN_Z-cut_GB_and_Qm}
\end{figure}

Although the floating structure provides valuable insight into the optical and mechanical modes of LNOI structures, its fabrications and thermal handling is impractical. When a substrate is added, as shown in \Cref{fig:LN_SBS_Waveguide}(b), the interaction is reshaped due to decreased light confinement and potential acoustic waves leakage based on the acoustic velocity contrast at the LN-substrate interface. The presence of a thick buried oxide layer also results in a high mechanical mode density, making it difficult to generate a useful dispersion diagram, like the one depicted in \Cref{fig:LN_Z-cut_Dispersions_and_GB}(b). we restricted our analysis to a fixed optical wavevector corresponding to the fundamental TE optical mode as a function of the waveguide rotation around the $Z$-axis, as shown in \Cref{fig:LN_Z-cut_GB_and_Qm}(a).

A key finding here is that even in the presence of a mechanically compliant substrate, mechanical surface waves (SAW) can be trapped in the LN device layer and provide a normalized Brillouin gain of 0.13 W$^{-1}$m$^{-1}$. Although it might be surprising that this mode has higher gain in the presence of a substrate, this occurs because the soft substrate partially clamps the bottom LN surface and creates larger vertical ($S_{yy}$) and longitudinal ($S_{zz}$) mechanical strains throughout the waveguide cross-section. Inspecting the mode profile of the lowest frequency mode, we can clearly correlate it to the mode labelled 3 in \Cref{fig:LN_Z-cut_Dispersions_and_GB}(c), but here the lower corners are attached to the substrate and the motion is concentrated in the vertical direction. Another important aspect is the complete mode confinement in the LN layer. Such feature of this mode is highlighted in \Cref{fig:LN_Z-cut_GB_and_Qm}(b), which shows the mechanical quality factor due to acoustic radiation towards the substrate. In this plot, the linewidth of each mode is chosen to match the mechanical linewidth ($\Gamma_{m}=\Omega_{m}/Q_{m}$), revealing that this is the only guided SAW mode in the $Z$-cut LNOI structure. Similarly to the suspended waveguide case, the optical stress density is dominated by the photoelastic contribution, in particular, by the $p_{14}$ and $S_{4}$ vertical-shear strain component.

\subsubsection{Ridge Waveguide $Z$-cut LN\label{sec:zridge}}

\begin{figure}[h!]
    \centering
    \includegraphics[width=7cm]{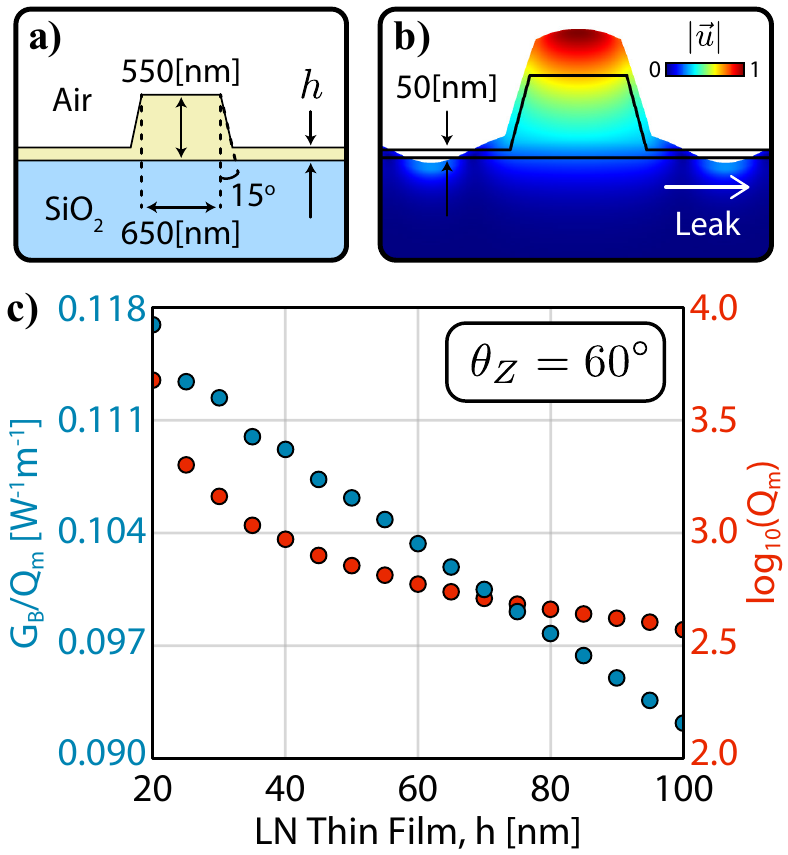}
    \caption{\textbf{a)} Ridge LN waveguide geometry parameters; \textbf{b)} Mechanical mode in analysis which, for the case of the strip waveguide, was a guided mode; \textbf{c)} Normalized Brillouin gain and mechanical quality factor as a function of the thickness $h$ for the fundamental mode at $\theta_{Z}=60^{\circ}$.}
    \label{fig:LN_Ridge-Z-cut_at_60degree_GB-and-Qm}
\end{figure}

Another common type of LNOI device is the ridge waveguide, which is characterized by a thin layer of LN is partially etched and left to enhance the interaction with electrodes in electro-optic modulators~\cite{Wang2018,Cortes-Herrera:21}, as depicted in \Cref{fig:LN_Ridge-Z-cut_at_60degree_GB-and-Qm}(a). 

\begin{figure*}[ht]
    \centering
    \includegraphics[width=16cm]{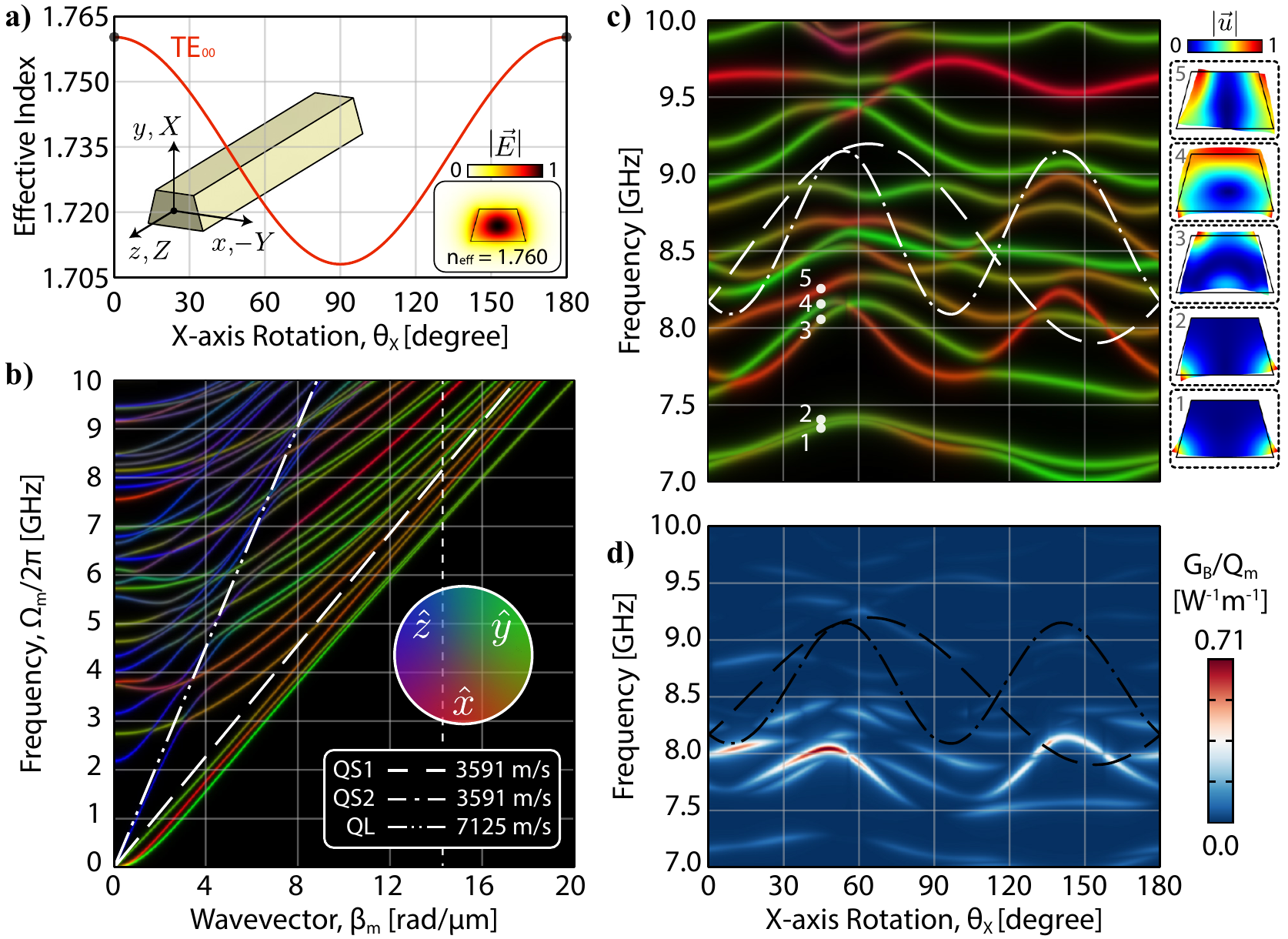}
    \caption{\textbf{a)} Fundamental optical TE mode effective refractive index at fixed $\lambda_{p}=1550$ nm as a function of $\theta_{X}$; \textbf{b)} $X$-cut mechanical dispersion analysis. The colors represent the same weighting of mechanical polarization as defined in \Cref{fig:LN_Z-cut_Dispersions_and_GB}. We used $\theta_{X}=0^{\circ}$ for this data; \textbf{c)} $X$-cut mechanical dispersion analysis as a function of $\theta_{X}$ with fixed $\beta_{m} = 14.27$ $\mu$m$^{-1}$; \textbf{d)} Normalized Brillouin gain density plot as a function of $\theta_{X}$. All data were simulated using the floating waveguide with $w=650$ nm, $t=550$ nm and $\phi=15^{\circ}$.}
    \label{fig:LN_X-cut_Dispersions_and_GB}
\end{figure*}

In this section, our aim is to evaluate the impact of this thin layer on the confinement of the large gain SAW modes, which were identified in \Cref{fig:LN_Z-cut_GB_and_Qm}. The presence of the LN layer leads to acoustic radiation leakage through the thin slab. This can be seen in the typical mode profile shown in \Cref{fig:LN_Ridge-Z-cut_at_60degree_GB-and-Qm}(b),which is consistence with the fact that the LN slab also supports guided SAW waves with similar dispersion properties to those of the LN strip waveguide top surface. Indeed, the radiation-limited mechanical quality factor suddenly drops for slab thicknesses $h>20$ nm, as shown in \Cref{fig:LN_Ridge-Z-cut_at_60degree_GB-and-Qm}(c); at $h=100$ nm, the radiation mechanical quality factor has fallen down to $Q_{m} \approx 300$, nevertheless, the modal spatial profile is not drastically affected, and the normalized gain is reduced only by approximately $20\%$. Thus, depending on the material-limited quality factors of LN, such structures may still be for exploring SBS in LNOI waveguides where the residual slab is required for electrode integration~\cite{Hu2021}.

\subsection{SBS in $X$-cut LN\label{sec:SBS_xcut}}

As discussed in \Cref{sec:anisotropy}, $X$-cut LN has a lower symmetry when compared to $Z$-cut orientation. Despite its lower symmetry, this cut is often explored due to its enhanced electro-optic and piezoelectric effects~\cite{Jiang:19}. \Cref{fig:LN_Slowness_and_PIJ}(f) also suggests that $X$-cut LN may have an enhanced SBS interaction due to the large photoelastic coefficients along specific directions. Therefore, this section analyzes the mechanical mode structure and SBS gain for the $X$-cut case, analogous to \Cref{sec:SBS_zcut}. Since the dielectric permittivity for this orientation is anisotropic, this is also considered when performing rotations around the crystal $X$-axis, ensuring that the optical mode takes this feature into account, as shown in \Cref{fig:LN_X-cut_Dispersions_and_GB}(a); note that although the optical field components vary when the crystal is rotated, the horizontal transverse polarization character is preserved. 

The mechanical dispersion relation at $\theta_{X}=0^{\circ}$ is shown in \Cref{fig:LN_X-cut_Dispersions_and_GB}(b) and has similar qualitative characteristics when compared with the $Z$-cut dispersion. However, we can clearly notice that the modes lying below the quasishear exhibit a larger degree of $x$-polarization, this is a consequence of the more anisotropic mechanical stiffness, as one could expect from the bulk velocity curves show in \Cref{fig:LN_Slowness_and_PIJ}(e). 
The low symmetry of this cut is fully revealed in \Cref{fig:LN_X-cut_Dispersions_and_GB}(c), where the phase-matched mechanical mode polarizations vary as the crystal is rotated around its $X$-axis. Not only the two-fold rotational symmetry is evident, but the mechanical modes show avoided crossing behaviors between $\hat{x}$ and $\hat{y}$ polarizations, as highlighted by the mode color changing. Nevertheless, the lowest frequency modes, identified in \Cref{fig:LN_X-cut_Dispersions_and_GB}(c), bare many similarities with those in \Cref{fig:LN_Z-cut_Dispersions_and_GB}(c).

The normalized Brillouin gain as a function of the $X$-axis rotation, $\theta_{X}$, is shown in \Cref{fig:LN_X-cut_Dispersions_and_GB}(d). As expected from the photoelastic angular dispersion, rather large normalized gains are observed around $\theta_{X} \approx 47^{\circ}$, reaching $G_{B}/Q_{m} \approx 0.71$ W$^{-1}$m$^{-1}$, almost three times larger than $Z$-cut. This angle is precisely where the $p_{12}$ is largest ($p_{12} \approx 0.28$), as we can verify at \Cref{fig:LN_Slowness_and_PIJ}(f). The individual contribution for each effect is shown in \Cref{supp} \Cref{fig:LN_PE_MB_RO}(d,e,f). If we compare the predictions by the simple plane-wave model of \Cref{eq:optimal_GB} for this mechanical mode using $\Omega_{m}/2\pi = 8$ GHz, $n_{\text{eff}} = 1.76$, $\beta_{m} = 14.27$ $\mu$m$^{-1}$, $p_{\text{eff}} = p_{13} \approx -0.1$ and $A_{\text{eff}} = 0.4$ $\mu$m$^{2}$, we get $G_{B}/Q_{m} \approx 0.087$ W$^{-1}$m$^{-1}$, which is one order of magnitude lower than the numerical prediction. This is expected because the strain in this mode is mostly vertically polarized, confirming that the anisotropy in LN should not be overlooked.

\subsubsection{Strip Waveguide $X$-cut LN\label{sec:xstrip}}

\begin{figure}[ht]
    \centering
    \includegraphics[width=7cm]{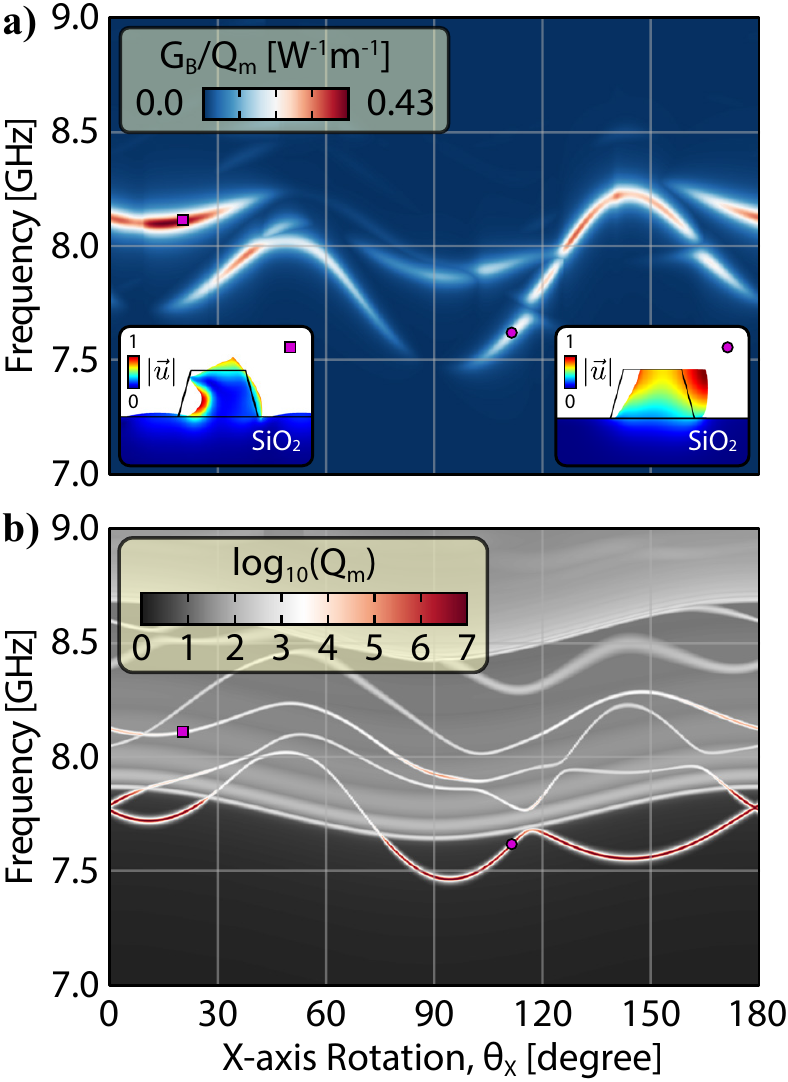}
    \caption{\textbf{a)} $X$-cut normalized Brillouin gain $G_{B}/Q_{m}$ density plot as a function of $\theta_{X}$. The purple square and the purple circle at $\theta_{X}=20^{\circ}$ and $\theta_{X}=110^{\circ}$, respectively, corresponds to the mechanical modes at the insets. The mechanical quality factor of the mode at $\theta_{X}=20^{\circ}$ is $Q_{m}=3450$ with an optical effective index of $n_{\text{eff}} = 1.7801$, while the one at $\theta_{X}=110^{\circ}$ is guided with effective index of $n_{\text{eff}} = 1.7414$; \textbf{b)} Mechanical quality factor $Q_{m}$ density plot as a function of $\theta_{X}$. We truncated log$_{10}\left(Q_{m}\right)$ at 7 to not saturate the map.}
    \label{fig:LN_X-cut_GB_and_Qm}
\end{figure}

Following the same approach explored for the $Z$-cut, the silicon dioxide substrate is considered in the simulations defining a ridge waveguide. Although the mode structure is more complex and the system has a lower symmetry compared to the $Z$-cut, acoustic guidance persists for specific angles, as shown in \Cref{fig:LN_X-cut_GB_and_Qm}.

Compared to the floating $X$-cut case, where we reached a maximum gain of 0.71 W$^{-1}$m$^{-1}$, the addition of the oxide substrate reduces the maximum gain to 0.43 W$^{-1}$m$^{-1}$, however, the topology of both \Cref{fig:LN_X-cut_Dispersions_and_GB}(d) and \Cref{fig:LN_X-cut_GB_and_Qm}(a) stays almost the same. Unfortunately, the maximum gain obtained isn't for a guided mode, but for a leaky mode, as shown in \Cref{fig:LN_X-cut_GB_and_Qm}(a) around 20$^{\circ}$ (left inset). \Cref{fig:LN_X-cut_GB_and_Qm}(b) shows the radiation-limited mechanical quality factor for this system, and we note that acoustic guidance occurs only at the angles between 0$^{\circ}$ and 30$^{\circ}$, as well for angles between 75$^{\circ}$ and 180$^{\circ}$, however, simultaneously guidance and large Brillouin gain was only achieved around 110$^{\circ}$ (roughly 0.2 W$^{-1}$m$^{-1}$), as shown at \Cref{fig:LN_X-cut_GB_and_Qm}(a) (right inset).

\subsubsection{Ridge Waveguide $X$-cut LN\label{sec:xridge}}

\begin{figure}[h!]
    \centering
    \includegraphics[width=7cm]{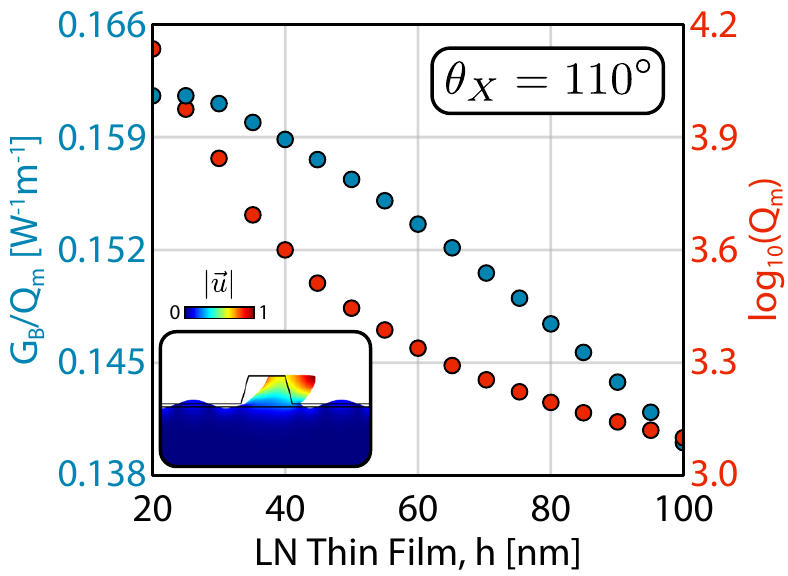}
    \caption{$X$-cut normalized Brillouin gain and mechanical quality factor as a function of the thickness $h$ for the fundamental mode at $\theta_{Z}=110^{\circ}$, which is shown at the inset for the case $h = 50$ nm.}
    \label{fig:LN_Ridge-X-cut_at_110degree_GB-and-Qm}
\end{figure}

The ridge waveguide for the $X$-cut suffers from the same problem as the $Z$-cut, i.e., the modes which were guided on the strip waveguide system become leaky modes. As shown in \Cref{fig:LN_Ridge-X-cut_at_110degree_GB-and-Qm}, there is a slight reduction in the SBS gain and a drop in the mechanical quality factor as residual LN slab thickens. The mode we choose to illustrate here corresponds to the one show in \Cref{fig:LN_X-cut_GB_and_Qm}(a) at $\theta_{X}=110^{\circ}$.

\newpage

\section{Conclusion}

A comprehensive numerical investigation of SBS in lithium niobate waveguides is presented, demonstrating that surface acoustic waves have a strong interaction with the optical fields. The most remarkable results are that for both $Z$-cut and $X$-cut orientations, SAW waves with high SBS gains were identified and guided in the LN surface, despite its higher longitudinal acoustic velocity when compared to the silicon dioxide substrate. The more symmetric $Z$-cut orientation offers a lower normalized Brillouin gain, $G_{B}/Q_{m}$, but has the potential to deliver material-limited mechanical quality factors. On the other hand, the low-symmetry $X$-cut orientation can exhibit significantly larger gain (3-fold enhancement compared to $Z$-cut), but the relevant mechanical modes exhibit substantial radiation loss. However, lower gain modes in the $X$-cut orientation (2-fold enhancement compared to $Z$-cut) can also be guided in the LN surface, but only at a narrow range of waveguide orientations. This is particularly critical when exploring SBS at low temperatures, where total mechanical quality factors have been reported to reach $1.7\times10^4$ at 2 GHz~\cite{Jiang:19} and $1.25\times10^4$ at 5.2 GHz~\cite{doi:10.1063/5.0020019} at cryogenic temperatures. These relatively high mechanical losses suggest that mechanical quality factors should not be limited by radiation, and values exceeding $10^3$ should be achievable at room temperature, as reported in similar geometries~\cite{doi:10.1063/5.0020019}. The main driver for SBS in LN is the photoelastic effect, with minor contributions from both roto-optic and moving boundary effects, which is a common feature in backward SBS~\cite{doi:10.1063/1.5088169}. Exploring the possibilities of SBS in LN could unleash new opportunities to study its second-order nonlinearities and piezoelectric properties, potentially revealing unprecedented interaction between optically excited high frequency mechanical waves and the radio-frequency domain. For instance, direct piezoelectric detection and excitation of optically pumped SBS waves could be explored for microwave photonics~\cite{Marpaung:15} or SBS lasing injection-locking applications~\cite{Spirin:06}. Further geometric optimization could also be considered to enhance the SBS gain in the proposed LNOI structures.\\

\noindent \textbf{Funding}. This work was supported by São Paulo Research Foundation (FAPESP) through grants
19/14377-5, %Caique 
22/11486-0,
19/13564-6, %Roberto
% 20/15786-3, %Pedro
% 19/09738-9, %Andre
% 17/19770-1, %Natalia 
% 19/01402-1, %Rodrigo
% 20/06348-2, %Caue
18/15577-5, %JP2-TA
18/15580-6, %JP2-GSW
18/25339-4, %Tematico Newton
% 22/07719-0, %TT-Toto
Coordenação de Aperfeiçoamento de Pessoal de Nível Superior - Brasil (CAPES) (Financial Code 001).\\

\noindent \textbf{Disclosures}.
The authors declare no conflicts of interest.\\

\noindent \textbf{Data availability}. All scripts files for generating each figure are available at Zenodo at  \href{http://doi.org/10.5281/zenodo.7314443}{10.5281/zenodo.7314443} upon publication.

\bibliography{main_refs.bib}

\begin{thebibliography}{52}%
\makeatletter
\providecommand \@ifxundefined [1]{%
 \@ifx{#1\undefined}
}%
\providecommand \@ifnum [1]{%
 \ifnum #1\expandafter \@firstoftwo
 \else \expandafter \@secondoftwo
 \fi
}%
\providecommand \@ifx [1]{%
 \ifx #1\expandafter \@firstoftwo
 \else \expandafter \@secondoftwo
 \fi
}%
\providecommand \natexlab [1]{#1}%
\providecommand \enquote  [1]{``#1''}%
\providecommand \bibnamefont  [1]{#1}%
\providecommand \bibfnamefont [1]{#1}%
\providecommand \citenamefont [1]{#1}%
\providecommand \href@noop [0]{\@secondoftwo}%
\providecommand \href [0]{\begingroup \@sanitize@url \@href}%
\providecommand \@href[1]{\@@startlink{#1}\@@href}%
\providecommand \@@href[1]{\endgroup#1\@@endlink}%
\providecommand \@sanitize@url [0]{\catcode `\\12\catcode `\$12\catcode
  `\&12\catcode `\#12\catcode `\^12\catcode `\_12\catcode `\%12\relax}%
\providecommand \@@startlink[1]{}%
\providecommand \@@endlink[0]{}%
\providecommand \url  [0]{\begingroup\@sanitize@url \@url }%
\providecommand \@url [1]{\endgroup\@href {#1}{\urlprefix }}%
\providecommand \urlprefix  [0]{URL }%
\providecommand \Eprint [0]{\href }%
\providecommand \doibase [0]{https://doi.org/}%
\providecommand \selectlanguage [0]{\@gobble}%
\providecommand \bibinfo  [0]{\@secondoftwo}%
\providecommand \bibfield  [0]{\@secondoftwo}%
\providecommand \translation [1]{[#1]}%
\providecommand \BibitemOpen [0]{}%
\providecommand \bibitemStop [0]{}%
\providecommand \bibitemNoStop [0]{.\EOS\space}%
\providecommand \EOS [0]{\spacefactor3000\relax}%
\providecommand \BibitemShut  [1]{\csname bibitem#1\endcsname}%
\let\auto@bib@innerbib\@empty
%</preamble>
\bibitem [{\citenamefont {Chiao}\ \emph {et~al.}(1964)\citenamefont {Chiao},
  \citenamefont {Townes},\ and\ \citenamefont
  {Stoicheff}}]{PhysRevLett.12.592}%
  \BibitemOpen
  \bibfield  {author} {\bibinfo {author} {\bibfnamefont {R.~Y.}\ \bibnamefont
  {Chiao}}, \bibinfo {author} {\bibfnamefont {C.~H.}\ \bibnamefont {Townes}},\
  and\ \bibinfo {author} {\bibfnamefont {B.~P.}\ \bibnamefont {Stoicheff}},\
  }\bibfield  {title} {\bibinfo {title} {Stimulated brillouin scattering and
  coherent generation of intense hypersonic waves},\ }\href
  {https://doi.org/10.1103/PhysRevLett.12.592} {\bibfield  {journal} {\bibinfo
  {journal} {Phys. Rev. Lett.}\ }\textbf {\bibinfo {volume} {12}},\ \bibinfo
  {pages} {592} (\bibinfo {year} {1964})}\BibitemShut {NoStop}%
\bibitem [{\citenamefont {Brewer}\ and\ \citenamefont
  {Rieckhoff}(1964)}]{PhysRevLett.13.334}%
  \BibitemOpen
  \bibfield  {author} {\bibinfo {author} {\bibfnamefont {R.~G.}\ \bibnamefont
  {Brewer}}\ and\ \bibinfo {author} {\bibfnamefont {K.~E.}\ \bibnamefont
  {Rieckhoff}},\ }\bibfield  {title} {\bibinfo {title} {Stimulated brillouin
  scattering in liquids},\ }\href {https://doi.org/10.1103/PhysRevLett.13.334}
  {\bibfield  {journal} {\bibinfo  {journal} {Phys. Rev. Lett.}\ }\textbf
  {\bibinfo {volume} {13}},\ \bibinfo {pages} {334} (\bibinfo {year}
  {1964})}\BibitemShut {NoStop}%
\bibitem [{\citenamefont {Gordon}(1968)}]{doi:10.1063/1.1655749}%
  \BibitemOpen
  \bibfield  {author} {\bibinfo {author} {\bibfnamefont {R.~L.}\ \bibnamefont
  {Gordon}},\ }\bibfield  {title} {\bibinfo {title} {Stimulated brillouin
  scattering in piezoelectric semiconductors},\ }\href
  {https://doi.org/10.1063/1.1655749} {\bibfield  {journal} {\bibinfo
  {journal} {Journal of Applied Physics}\ }\textbf {\bibinfo {volume} {39}},\
  \bibinfo {pages} {306} (\bibinfo {year} {1968})}\BibitemShut {NoStop}%
\bibitem [{\citenamefont {Sen}\ and\ \citenamefont
  {Sen}(1985)}]{PhysRevB.31.1034}%
  \BibitemOpen
  \bibfield  {author} {\bibinfo {author} {\bibfnamefont {P.}~\bibnamefont
  {Sen}}\ and\ \bibinfo {author} {\bibfnamefont {P.~K.}\ \bibnamefont {Sen}},\
  }\bibfield  {title} {\bibinfo {title} {Theory of stimulated raman and
  brillouin scattering in noncentrosymmetric crystals},\ }\href
  {https://doi.org/10.1103/PhysRevB.31.1034} {\bibfield  {journal} {\bibinfo
  {journal} {Phys. Rev. B}\ }\textbf {\bibinfo {volume} {31}},\ \bibinfo
  {pages} {1034} (\bibinfo {year} {1985})}\BibitemShut {NoStop}%
\bibitem [{\citenamefont {Ippen}\ and\ \citenamefont
  {Stolen}(1972)}]{doi:10.1063/1.1654249}%
  \BibitemOpen
  \bibfield  {author} {\bibinfo {author} {\bibfnamefont {E.}~\bibnamefont
  {Ippen}}\ and\ \bibinfo {author} {\bibfnamefont {R.}~\bibnamefont {Stolen}},\
  }\bibfield  {title} {\bibinfo {title} {Stimulated brillouin scattering in
  optical fibers},\ }\href {https://doi.org/10.1063/1.1654249} {\bibfield
  {journal} {\bibinfo  {journal} {Applied Physics Letters}\ }\textbf {\bibinfo
  {volume} {21}},\ \bibinfo {pages} {539} (\bibinfo {year} {1972})}\BibitemShut
  {NoStop}%
\bibitem [{\citenamefont {Hill}\ \emph {et~al.}(1976)\citenamefont {Hill},
  \citenamefont {Kawasaki},\ and\ \citenamefont
  {Johnson}}]{doi:10.1063/1.88583}%
  \BibitemOpen
  \bibfield  {author} {\bibinfo {author} {\bibfnamefont {K.~O.}\ \bibnamefont
  {Hill}}, \bibinfo {author} {\bibfnamefont {B.~S.}\ \bibnamefont {Kawasaki}},\
  and\ \bibinfo {author} {\bibfnamefont {D.~C.}\ \bibnamefont {Johnson}},\
  }\bibfield  {title} {\bibinfo {title} {cw brillouin laser},\ }\href
  {https://doi.org/10.1063/1.88583} {\bibfield  {journal} {\bibinfo  {journal}
  {Applied Physics Letters}\ }\textbf {\bibinfo {volume} {28}},\ \bibinfo
  {pages} {608} (\bibinfo {year} {1976})}\BibitemShut {NoStop}%
\bibitem [{\citenamefont {Rowell}\ \emph {et~al.}(1979)\citenamefont {Rowell},
  \citenamefont {Thomas}, \citenamefont {van Driel},\ and\ \citenamefont
  {Stegeman}}]{doi:10.1063/1.90726}%
  \BibitemOpen
  \bibfield  {author} {\bibinfo {author} {\bibfnamefont {N.~L.}\ \bibnamefont
  {Rowell}}, \bibinfo {author} {\bibfnamefont {P.~J.}\ \bibnamefont {Thomas}},
  \bibinfo {author} {\bibfnamefont {H.~M.}\ \bibnamefont {van Driel}},\ and\
  \bibinfo {author} {\bibfnamefont {G.~I.}\ \bibnamefont {Stegeman}},\
  }\bibfield  {title} {\bibinfo {title} {Brillouin spectrum of single‐mode
  optical fibers},\ }\href {https://doi.org/10.1063/1.90726} {\bibfield
  {journal} {\bibinfo  {journal} {Applied Physics Letters}\ }\textbf {\bibinfo
  {volume} {34}},\ \bibinfo {pages} {139} (\bibinfo {year} {1979})}\BibitemShut
  {NoStop}%
\bibitem [{\citenamefont {Rowell}\ \emph {et~al.}(1978)\citenamefont {Rowell},
  \citenamefont {So},\ and\ \citenamefont {Stegeman}}]{doi:10.1063/1.89965}%
  \BibitemOpen
  \bibfield  {author} {\bibinfo {author} {\bibfnamefont {N.}~\bibnamefont
  {Rowell}}, \bibinfo {author} {\bibfnamefont {V.~C.}\ \bibnamefont {So}},\
  and\ \bibinfo {author} {\bibfnamefont {G.~I.}\ \bibnamefont {Stegeman}},\
  }\bibfield  {title} {\bibinfo {title} {Brillouin scattering in a thin‐film
  waveguide},\ }\href {https://doi.org/10.1063/1.89965} {\bibfield  {journal}
  {\bibinfo  {journal} {Applied Physics Letters}\ }\textbf {\bibinfo {volume}
  {32}},\ \bibinfo {pages} {154} (\bibinfo {year} {1978})}\BibitemShut
  {NoStop}%
\bibitem [{\citenamefont {Pant}\ \emph {et~al.}(2011)\citenamefont {Pant},
  \citenamefont {Poulton}, \citenamefont {Choi}, \citenamefont {Mcfarlane},
  \citenamefont {Hile}, \citenamefont {Li}, \citenamefont {Thevenaz},
  \citenamefont {Luther-Davies}, \citenamefont {Madden},\ and\ \citenamefont
  {Eggleton}}]{Pant:11}%
  \BibitemOpen
  \bibfield  {author} {\bibinfo {author} {\bibfnamefont {R.}~\bibnamefont
  {Pant}}, \bibinfo {author} {\bibfnamefont {C.~G.}\ \bibnamefont {Poulton}},
  \bibinfo {author} {\bibfnamefont {D.-Y.}\ \bibnamefont {Choi}}, \bibinfo
  {author} {\bibfnamefont {H.}~\bibnamefont {Mcfarlane}}, \bibinfo {author}
  {\bibfnamefont {S.}~\bibnamefont {Hile}}, \bibinfo {author} {\bibfnamefont
  {E.}~\bibnamefont {Li}}, \bibinfo {author} {\bibfnamefont {L.}~\bibnamefont
  {Thevenaz}}, \bibinfo {author} {\bibfnamefont {B.}~\bibnamefont
  {Luther-Davies}}, \bibinfo {author} {\bibfnamefont {S.~J.}\ \bibnamefont
  {Madden}},\ and\ \bibinfo {author} {\bibfnamefont {B.~J.}\ \bibnamefont
  {Eggleton}},\ }\bibfield  {title} {\bibinfo {title} {On-chip stimulated
  brillouin scattering},\ }\href {https://doi.org/10.1364/OE.19.008285}
  {\bibfield  {journal} {\bibinfo  {journal} {Opt. Express}\ }\textbf {\bibinfo
  {volume} {19}},\ \bibinfo {pages} {8285} (\bibinfo {year}
  {2011})}\BibitemShut {NoStop}%
\bibitem [{\citenamefont {Shin}\ \emph {et~al.}(2013)\citenamefont {Shin},
  \citenamefont {Qiu}, \citenamefont {Jarecki}, \citenamefont {Cox},
  \citenamefont {Olsson}, \citenamefont {Starbuck}, \citenamefont {Wang},\ and\
  \citenamefont {Rakich}}]{Shin2013}%
  \BibitemOpen
  \bibfield  {author} {\bibinfo {author} {\bibfnamefont {H.}~\bibnamefont
  {Shin}}, \bibinfo {author} {\bibfnamefont {W.}~\bibnamefont {Qiu}}, \bibinfo
  {author} {\bibfnamefont {R.}~\bibnamefont {Jarecki}}, \bibinfo {author}
  {\bibfnamefont {J.~A.}\ \bibnamefont {Cox}}, \bibinfo {author} {\bibfnamefont
  {R.~H.}\ \bibnamefont {Olsson}}, \bibinfo {author} {\bibfnamefont
  {A.}~\bibnamefont {Starbuck}}, \bibinfo {author} {\bibfnamefont
  {Z.}~\bibnamefont {Wang}},\ and\ \bibinfo {author} {\bibfnamefont {P.~T.}\
  \bibnamefont {Rakich}},\ }\bibfield  {title} {\bibinfo {title} {Tailorable
  stimulated brillouin scattering in nanoscale silicon waveguides},\ }\href
  {https://doi.org/10.1038/ncomms2943} {\bibfield  {journal} {\bibinfo
  {journal} {Nature Communications}\ }\textbf {\bibinfo {volume} {4}},\
  \bibinfo {pages} {1944} (\bibinfo {year} {2013})}\BibitemShut {NoStop}%
\bibitem [{\citenamefont {Van~Laer}\ \emph {et~al.}(2015)\citenamefont
  {Van~Laer}, \citenamefont {Kuyken}, \citenamefont {Van~Thourhout},\ and\
  \citenamefont {Baets}}]{VanLaer2015}%
  \BibitemOpen
  \bibfield  {author} {\bibinfo {author} {\bibfnamefont {R.}~\bibnamefont
  {Van~Laer}}, \bibinfo {author} {\bibfnamefont {B.}~\bibnamefont {Kuyken}},
  \bibinfo {author} {\bibfnamefont {D.}~\bibnamefont {Van~Thourhout}},\ and\
  \bibinfo {author} {\bibfnamefont {R.}~\bibnamefont {Baets}},\ }\bibfield
  {title} {\bibinfo {title} {Interaction between light and highly confined
  hypersound in a silicon photonic nanowire},\ }\href
  {https://doi.org/10.1038/nphoton.2015.11} {\bibfield  {journal} {\bibinfo
  {journal} {Nature Photonics}\ }\textbf {\bibinfo {volume} {9}},\ \bibinfo
  {pages} {199} (\bibinfo {year} {2015})}\BibitemShut {NoStop}%
\bibitem [{\citenamefont {Gyger}\ \emph {et~al.}(2020)\citenamefont {Gyger},
  \citenamefont {Liu}, \citenamefont {Yang}, \citenamefont {He}, \citenamefont
  {Raja}, \citenamefont {Wang}, \citenamefont {Bhave}, \citenamefont
  {Kippenberg},\ and\ \citenamefont {Th\'evenaz}}]{PhysRevLett.124.013902}%
  \BibitemOpen
  \bibfield  {author} {\bibinfo {author} {\bibfnamefont {F.}~\bibnamefont
  {Gyger}}, \bibinfo {author} {\bibfnamefont {J.}~\bibnamefont {Liu}}, \bibinfo
  {author} {\bibfnamefont {F.}~\bibnamefont {Yang}}, \bibinfo {author}
  {\bibfnamefont {J.}~\bibnamefont {He}}, \bibinfo {author} {\bibfnamefont
  {A.~S.}\ \bibnamefont {Raja}}, \bibinfo {author} {\bibfnamefont {R.~N.}\
  \bibnamefont {Wang}}, \bibinfo {author} {\bibfnamefont {S.~A.}\ \bibnamefont
  {Bhave}}, \bibinfo {author} {\bibfnamefont {T.~J.}\ \bibnamefont
  {Kippenberg}},\ and\ \bibinfo {author} {\bibfnamefont {L.}~\bibnamefont
  {Th\'evenaz}},\ }\bibfield  {title} {\bibinfo {title} {Observation of
  stimulated brillouin scattering in silicon nitride integrated waveguides},\
  }\href {https://doi.org/10.1103/PhysRevLett.124.013902} {\bibfield  {journal}
  {\bibinfo  {journal} {Phys. Rev. Lett.}\ }\textbf {\bibinfo {volume} {124}},\
  \bibinfo {pages} {013902} (\bibinfo {year} {2020})}\BibitemShut {NoStop}%
\bibitem [{\citenamefont {Botter}\ \emph {et~al.}(2022)\citenamefont {Botter},
  \citenamefont {Ye}, \citenamefont {Klaver}, \citenamefont {Suryadharma},
  \citenamefont {Daulay}, \citenamefont {Liu}, \citenamefont {van~den Hoogen},
  \citenamefont {Kanger}, \citenamefont {van~der Slot}, \citenamefont {Klein},
  \citenamefont {Hoekman}, \citenamefont {Roeloffzen}, \citenamefont {Liu},\
  and\ \citenamefont {Marpaung}}]{doi:10.1126/sciadv.abq2196}%
  \BibitemOpen
  \bibfield  {author} {\bibinfo {author} {\bibfnamefont {R.}~\bibnamefont
  {Botter}}, \bibinfo {author} {\bibfnamefont {K.}~\bibnamefont {Ye}}, \bibinfo
  {author} {\bibfnamefont {Y.}~\bibnamefont {Klaver}}, \bibinfo {author}
  {\bibfnamefont {R.}~\bibnamefont {Suryadharma}}, \bibinfo {author}
  {\bibfnamefont {O.}~\bibnamefont {Daulay}}, \bibinfo {author} {\bibfnamefont
  {G.}~\bibnamefont {Liu}}, \bibinfo {author} {\bibfnamefont {J.}~\bibnamefont
  {van~den Hoogen}}, \bibinfo {author} {\bibfnamefont {L.}~\bibnamefont
  {Kanger}}, \bibinfo {author} {\bibfnamefont {P.}~\bibnamefont {van~der
  Slot}}, \bibinfo {author} {\bibfnamefont {E.}~\bibnamefont {Klein}}, \bibinfo
  {author} {\bibfnamefont {M.}~\bibnamefont {Hoekman}}, \bibinfo {author}
  {\bibfnamefont {C.}~\bibnamefont {Roeloffzen}}, \bibinfo {author}
  {\bibfnamefont {Y.}~\bibnamefont {Liu}},\ and\ \bibinfo {author}
  {\bibfnamefont {D.}~\bibnamefont {Marpaung}},\ }\bibfield  {title} {\bibinfo
  {title} {Guided-acoustic stimulated brillouin scattering in silicon nitride
  photonic circuits},\ }\href {https://doi.org/10.1126/sciadv.abq2196}
  {\bibfield  {journal} {\bibinfo  {journal} {Science Advances}\ }\textbf
  {\bibinfo {volume} {8}},\ \bibinfo {pages} {eabq2196} (\bibinfo {year}
  {2022})}\BibitemShut {NoStop}%
\bibitem [{\citenamefont {Jiang}\ and\ \citenamefont {Lin}(2016)}]{Jiang2016}%
  \BibitemOpen
  \bibfield  {author} {\bibinfo {author} {\bibfnamefont {W.~C.}\ \bibnamefont
  {Jiang}}\ and\ \bibinfo {author} {\bibfnamefont {Q.}~\bibnamefont {Lin}},\
  }\bibfield  {title} {\bibinfo {title} {Chip-scale cavity optomechanics in
  lithium niobate},\ }\href {https://doi.org/10.1038/srep36920} {\bibfield
  {journal} {\bibinfo  {journal} {Scientific Reports}\ }\textbf {\bibinfo
  {volume} {6}},\ \bibinfo {pages} {36920} (\bibinfo {year}
  {2016})}\BibitemShut {NoStop}%
\bibitem [{\citenamefont {Jiang}\ \emph {et~al.}(2019)\citenamefont {Jiang},
  \citenamefont {Patel}, \citenamefont {Mayor}, \citenamefont {McKenna},
  \citenamefont {Arrangoiz-Arriola}, \citenamefont {Sarabalis}, \citenamefont
  {Witmer}, \citenamefont {Laer},\ and\ \citenamefont
  {Safavi-Naeini}}]{Jiang:19}%
  \BibitemOpen
  \bibfield  {author} {\bibinfo {author} {\bibfnamefont {W.}~\bibnamefont
  {Jiang}}, \bibinfo {author} {\bibfnamefont {R.~N.}\ \bibnamefont {Patel}},
  \bibinfo {author} {\bibfnamefont {F.~M.}\ \bibnamefont {Mayor}}, \bibinfo
  {author} {\bibfnamefont {T.~P.}\ \bibnamefont {McKenna}}, \bibinfo {author}
  {\bibfnamefont {P.}~\bibnamefont {Arrangoiz-Arriola}}, \bibinfo {author}
  {\bibfnamefont {C.~J.}\ \bibnamefont {Sarabalis}}, \bibinfo {author}
  {\bibfnamefont {J.~D.}\ \bibnamefont {Witmer}}, \bibinfo {author}
  {\bibfnamefont {R.~V.}\ \bibnamefont {Laer}},\ and\ \bibinfo {author}
  {\bibfnamefont {A.~H.}\ \bibnamefont {Safavi-Naeini}},\ }\bibfield  {title}
  {\bibinfo {title} {Lithium niobate piezo-optomechanical crystals},\ }\href
  {https://doi.org/10.1364/OPTICA.6.000845} {\bibfield  {journal} {\bibinfo
  {journal} {Optica}\ }\textbf {\bibinfo {volume} {6}},\ \bibinfo {pages} {845}
  (\bibinfo {year} {2019})}\BibitemShut {NoStop}%
\bibitem [{\citenamefont {Örsel}\ and\ \citenamefont
  {Bahl}(2022)}]{https://doi.org/10.48550/arxiv.2210.01064}%
  \BibitemOpen
  \bibfield  {author} {\bibinfo {author} {\bibfnamefont {O.~E.}\ \bibnamefont
  {Örsel}}\ and\ \bibinfo {author} {\bibfnamefont {G.}~\bibnamefont {Bahl}},\
  }\href {https://doi.org/10.48550/ARXIV.2210.01064} {\bibinfo {title}
  {Electro-optic non-reciprocal polarization rotation in lithium niobate}}
  (\bibinfo {year} {2022})\BibitemShut {NoStop}%
\bibitem [{\citenamefont {Wolff}\ \emph {et~al.}(2015)\citenamefont {Wolff},
  \citenamefont {Steel}, \citenamefont {Eggleton},\ and\ \citenamefont
  {Poulton}}]{PhysRevA.92.013836}%
  \BibitemOpen
  \bibfield  {author} {\bibinfo {author} {\bibfnamefont {C.}~\bibnamefont
  {Wolff}}, \bibinfo {author} {\bibfnamefont {M.~J.}\ \bibnamefont {Steel}},
  \bibinfo {author} {\bibfnamefont {B.~J.}\ \bibnamefont {Eggleton}},\ and\
  \bibinfo {author} {\bibfnamefont {C.~G.}\ \bibnamefont {Poulton}},\
  }\bibfield  {title} {\bibinfo {title} {Stimulated brillouin scattering in
  integrated photonic waveguides: Forces, scattering mechanisms, and
  coupled-mode analysis},\ }\href {https://doi.org/10.1103/PhysRevA.92.013836}
  {\bibfield  {journal} {\bibinfo  {journal} {Phys. Rev. A}\ }\textbf {\bibinfo
  {volume} {92}},\ \bibinfo {pages} {013836} (\bibinfo {year}
  {2015})}\BibitemShut {NoStop}%
\bibitem [{\citenamefont {Eggleton}\ \emph {et~al.}(2019)\citenamefont
  {Eggleton}, \citenamefont {Poulton}, \citenamefont {Rakich}, \citenamefont
  {Steel},\ and\ \citenamefont {Bahl}}]{Eggleton2019}%
  \BibitemOpen
  \bibfield  {author} {\bibinfo {author} {\bibfnamefont {B.~J.}\ \bibnamefont
  {Eggleton}}, \bibinfo {author} {\bibfnamefont {C.~G.}\ \bibnamefont
  {Poulton}}, \bibinfo {author} {\bibfnamefont {P.~T.}\ \bibnamefont {Rakich}},
  \bibinfo {author} {\bibfnamefont {M.~J.}\ \bibnamefont {Steel}},\ and\
  \bibinfo {author} {\bibfnamefont {G.}~\bibnamefont {Bahl}},\ }\bibfield
  {title} {\bibinfo {title} {Brillouin integrated photonics},\ }\href
  {https://doi.org/10.1038/s41566-019-0498-z} {\bibfield  {journal} {\bibinfo
  {journal} {Nature Photonics}\ }\textbf {\bibinfo {volume} {13}},\ \bibinfo
  {pages} {664} (\bibinfo {year} {2019})}\BibitemShut {NoStop}%
\bibitem [{\citenamefont {Poulton}\ \emph {et~al.}(2013)\citenamefont
  {Poulton}, \citenamefont {Pant},\ and\ \citenamefont
  {Eggleton}}]{Poulton:13}%
  \BibitemOpen
  \bibfield  {author} {\bibinfo {author} {\bibfnamefont {C.~G.}\ \bibnamefont
  {Poulton}}, \bibinfo {author} {\bibfnamefont {R.}~\bibnamefont {Pant}},\ and\
  \bibinfo {author} {\bibfnamefont {B.~J.}\ \bibnamefont {Eggleton}},\
  }\bibfield  {title} {\bibinfo {title} {Acoustic confinement and stimulated
  brillouin scattering in integrated optical waveguides},\ }\href
  {https://doi.org/10.1364/JOSAB.30.002657} {\bibfield  {journal} {\bibinfo
  {journal} {J. Opt. Soc. Am. B}\ }\textbf {\bibinfo {volume} {30}},\ \bibinfo
  {pages} {2657} (\bibinfo {year} {2013})}\BibitemShut {NoStop}%
\bibitem [{\citenamefont {Eggleton}\ \emph {et~al.}(2013)\citenamefont
  {Eggleton}, \citenamefont {Poulton},\ and\ \citenamefont
  {Pant}}]{Eggleton:13}%
  \BibitemOpen
  \bibfield  {author} {\bibinfo {author} {\bibfnamefont {B.~J.}\ \bibnamefont
  {Eggleton}}, \bibinfo {author} {\bibfnamefont {C.~G.}\ \bibnamefont
  {Poulton}},\ and\ \bibinfo {author} {\bibfnamefont {R.}~\bibnamefont
  {Pant}},\ }\bibfield  {title} {\bibinfo {title} {Inducing and harnessing
  stimulated brillouin scattering in photonic integrated circuits},\ }\href
  {https://doi.org/10.1364/AOP.5.000536} {\bibfield  {journal} {\bibinfo
  {journal} {Adv. Opt. Photon.}\ }\textbf {\bibinfo {volume} {5}},\ \bibinfo
  {pages} {536} (\bibinfo {year} {2013})}\BibitemShut {NoStop}%
\bibitem [{\citenamefont {Van~Laer}\ \emph {et~al.}(2016)\citenamefont
  {Van~Laer}, \citenamefont {Baets},\ and\ \citenamefont
  {Van~Thourhout}}]{PhysRevA.93.053828}%
  \BibitemOpen
  \bibfield  {author} {\bibinfo {author} {\bibfnamefont {R.}~\bibnamefont
  {Van~Laer}}, \bibinfo {author} {\bibfnamefont {R.}~\bibnamefont {Baets}},\
  and\ \bibinfo {author} {\bibfnamefont {D.}~\bibnamefont {Van~Thourhout}},\
  }\bibfield  {title} {\bibinfo {title} {Unifying brillouin scattering and
  cavity optomechanics},\ }\href {https://doi.org/10.1103/PhysRevA.93.053828}
  {\bibfield  {journal} {\bibinfo  {journal} {Phys. Rev. A}\ }\textbf {\bibinfo
  {volume} {93}},\ \bibinfo {pages} {053828} (\bibinfo {year}
  {2016})}\BibitemShut {NoStop}%
\bibitem [{\citenamefont {Sipe}\ and\ \citenamefont
  {Steel}(2016)}]{JESipe_2016}%
  \BibitemOpen
  \bibfield  {author} {\bibinfo {author} {\bibfnamefont {J.~E.}\ \bibnamefont
  {Sipe}}\ and\ \bibinfo {author} {\bibfnamefont {M.~J.}\ \bibnamefont
  {Steel}},\ }\bibfield  {title} {\bibinfo {title} {A hamiltonian treatment of
  stimulated brillouin scattering in nanoscale integrated waveguides},\ }\href
  {https://doi.org/10.1088/1367-2630/18/4/045004} {\bibfield  {journal}
  {\bibinfo  {journal} {New Journal of Physics}\ }\textbf {\bibinfo {volume}
  {18}},\ \bibinfo {pages} {045004} (\bibinfo {year} {2016})}\BibitemShut
  {NoStop}%
\bibitem [{\citenamefont {Wolff}\ \emph {et~al.}(2021)\citenamefont {Wolff},
  \citenamefont {Smith}, \citenamefont {Stiller},\ and\ \citenamefont
  {Poulton}}]{Wolff:21}%
  \BibitemOpen
  \bibfield  {author} {\bibinfo {author} {\bibfnamefont {C.}~\bibnamefont
  {Wolff}}, \bibinfo {author} {\bibfnamefont {M.~J.~A.}\ \bibnamefont {Smith}},
  \bibinfo {author} {\bibfnamefont {B.}~\bibnamefont {Stiller}},\ and\ \bibinfo
  {author} {\bibfnamefont {C.~G.}\ \bibnamefont {Poulton}},\ }\bibfield
  {title} {\bibinfo {title} {Brillouin scattering---theory and experiment:
  tutorial},\ }\href {https://doi.org/10.1364/JOSAB.416747} {\bibfield
  {journal} {\bibinfo  {journal} {J. Opt. Soc. Am. B}\ }\textbf {\bibinfo
  {volume} {38}},\ \bibinfo {pages} {1243} (\bibinfo {year}
  {2021})}\BibitemShut {NoStop}%
\bibitem [{\citenamefont {Wiederhecker}\ \emph {et~al.}(2019)\citenamefont
  {Wiederhecker}, \citenamefont {Dainese},\ and\ \citenamefont
  {Mayer~Alegre}}]{doi:10.1063/1.5088169}%
  \BibitemOpen
  \bibfield  {author} {\bibinfo {author} {\bibfnamefont {G.~S.}\ \bibnamefont
  {Wiederhecker}}, \bibinfo {author} {\bibfnamefont {P.}~\bibnamefont
  {Dainese}},\ and\ \bibinfo {author} {\bibfnamefont {T.~P.}\ \bibnamefont
  {Mayer~Alegre}},\ }\bibfield  {title} {\bibinfo {title} {Brillouin
  optomechanics in nanophotonic structures},\ }\href
  {https://doi.org/10.1063/1.5088169} {\bibfield  {journal} {\bibinfo
  {journal} {APL Photonics}\ }\textbf {\bibinfo {volume} {4}},\ \bibinfo
  {pages} {071101} (\bibinfo {year} {2019})}\BibitemShut {NoStop}%
\bibitem [{\citenamefont {Wang}\ \emph {et~al.}(2021)\citenamefont {Wang},
  \citenamefont {Yu}, \citenamefont {Li}, \citenamefont {Bai}, \citenamefont
  {Wang}, \citenamefont {Li}, \citenamefont {Song}, \citenamefont {Wang},
  \citenamefont {Li}, \citenamefont {Wang}, \citenamefont {Lu}, \citenamefont
  {Li}, \citenamefont {Liu},\ and\ \citenamefont {Yan}}]{app11188390}%
  \BibitemOpen
  \bibfield  {author} {\bibinfo {author} {\bibfnamefont {W.}~\bibnamefont
  {Wang}}, \bibinfo {author} {\bibfnamefont {Y.}~\bibnamefont {Yu}}, \bibinfo
  {author} {\bibfnamefont {Y.}~\bibnamefont {Li}}, \bibinfo {author}
  {\bibfnamefont {Z.}~\bibnamefont {Bai}}, \bibinfo {author} {\bibfnamefont
  {G.}~\bibnamefont {Wang}}, \bibinfo {author} {\bibfnamefont {K.}~\bibnamefont
  {Li}}, \bibinfo {author} {\bibfnamefont {C.}~\bibnamefont {Song}}, \bibinfo
  {author} {\bibfnamefont {Z.}~\bibnamefont {Wang}}, \bibinfo {author}
  {\bibfnamefont {S.}~\bibnamefont {Li}}, \bibinfo {author} {\bibfnamefont
  {Y.}~\bibnamefont {Wang}}, \bibinfo {author} {\bibfnamefont {Z.}~\bibnamefont
  {Lu}}, \bibinfo {author} {\bibfnamefont {Y.}~\bibnamefont {Li}}, \bibinfo
  {author} {\bibfnamefont {T.}~\bibnamefont {Liu}},\ and\ \bibinfo {author}
  {\bibfnamefont {X.}~\bibnamefont {Yan}},\ }\bibfield  {title} {\bibinfo
  {title} {Tailorable brillouin light scattering in a lithium niobate
  waveguide},\ }\bibfield  {journal} {\bibinfo  {journal} {Applied Sciences}\
  }\textbf {\bibinfo {volume} {11}},\ \href
  {https://doi.org/10.3390/app11188390} {10.3390/app11188390} (\bibinfo {year}
  {2021})\BibitemShut {NoStop}%
\bibitem [{\citenamefont {Weis}\ and\ \citenamefont
  {Gaylord}(1985)}]{Weis1985}%
  \BibitemOpen
  \bibfield  {author} {\bibinfo {author} {\bibfnamefont {R.~S.}\ \bibnamefont
  {Weis}}\ and\ \bibinfo {author} {\bibfnamefont {T.~K.}\ \bibnamefont
  {Gaylord}},\ }\bibfield  {title} {\bibinfo {title} {Lithium niobate: Summary
  of physical properties and crystal structure},\ }\href
  {https://doi.org/10.1007/BF00614817} {\bibfield  {journal} {\bibinfo
  {journal} {Applied Physics A}\ }\textbf {\bibinfo {volume} {37}},\ \bibinfo
  {pages} {191} (\bibinfo {year} {1985})}\BibitemShut {NoStop}%
\bibitem [{\citenamefont {Abrahams}\ \emph {et~al.}(1966)\citenamefont
  {Abrahams}, \citenamefont {Reddy},\ and\ \citenamefont
  {Bernstein}}]{ABRAHAMS1966997}%
  \BibitemOpen
  \bibfield  {author} {\bibinfo {author} {\bibfnamefont {S.}~\bibnamefont
  {Abrahams}}, \bibinfo {author} {\bibfnamefont {J.}~\bibnamefont {Reddy}},\
  and\ \bibinfo {author} {\bibfnamefont {J.}~\bibnamefont {Bernstein}},\
  }\bibfield  {title} {\bibinfo {title} {Ferroelectric lithium niobate. 3.
  single crystal x-ray diffraction study at 24°c},\ }\href
  {https://doi.org/https://doi.org/10.1016/0022-3697(66)90072-2} {\bibfield
  {journal} {\bibinfo  {journal} {Journal of Physics and Chemistry of Solids}\
  }\textbf {\bibinfo {volume} {27}},\ \bibinfo {pages} {997} (\bibinfo {year}
  {1966})}\BibitemShut {NoStop}%
\bibitem [{\citenamefont {Smith}\ and\ \citenamefont
  {Welsh}(1971)}]{doi:10.1063/1.1660528}%
  \BibitemOpen
  \bibfield  {author} {\bibinfo {author} {\bibfnamefont {R.~T.}\ \bibnamefont
  {Smith}}\ and\ \bibinfo {author} {\bibfnamefont {F.~S.}\ \bibnamefont
  {Welsh}},\ }\bibfield  {title} {\bibinfo {title} {Temperature dependence of
  the elastic, piezoelectric, and dielectric constants of lithium tantalate and
  lithium niobate},\ }\href {https://doi.org/10.1063/1.1660528} {\bibfield
  {journal} {\bibinfo  {journal} {Journal of Applied Physics}\ }\textbf
  {\bibinfo {volume} {42}},\ \bibinfo {pages} {2219} (\bibinfo {year}
  {1971})}\BibitemShut {NoStop}%
\bibitem [{\citenamefont {Yamada}\ \emph {et~al.}(1967)\citenamefont {Yamada},
  \citenamefont {Niizeki},\ and\ \citenamefont {Toyoda}}]{Yamada_1967}%
  \BibitemOpen
  \bibfield  {author} {\bibinfo {author} {\bibfnamefont {T.}~\bibnamefont
  {Yamada}}, \bibinfo {author} {\bibfnamefont {N.}~\bibnamefont {Niizeki}},\
  and\ \bibinfo {author} {\bibfnamefont {H.}~\bibnamefont {Toyoda}},\
  }\bibfield  {title} {\bibinfo {title} {Piezoelectric and elastic properties
  of lithium niobate single crystals},\ }\href
  {https://doi.org/10.1143/jjap.6.151} {\bibfield  {journal} {\bibinfo
  {journal} {Japanese Journal of Applied Physics}\ }\textbf {\bibinfo {volume}
  {6}},\ \bibinfo {pages} {151} (\bibinfo {year} {1967})}\BibitemShut {NoStop}%
\bibitem [{\citenamefont {Jazbin{\v{s}}ek}\ and\ \citenamefont
  {Zgonik}(2002)}]{Jazbinsek2002}%
  \BibitemOpen
  \bibfield  {author} {\bibinfo {author} {\bibfnamefont {M.}~\bibnamefont
  {Jazbin{\v{s}}ek}}\ and\ \bibinfo {author} {\bibfnamefont {M.}~\bibnamefont
  {Zgonik}},\ }\bibfield  {title} {\bibinfo {title} {Material tensor parameters
  of linbo3 relevant for electro- and elasto-optics},\ }\href
  {https://doi.org/10.1007/s003400200818} {\bibfield  {journal} {\bibinfo
  {journal} {Applied Physics B}\ }\textbf {\bibinfo {volume} {74}},\ \bibinfo
  {pages} {407} (\bibinfo {year} {2002})}\BibitemShut {NoStop}%
\bibitem [{\citenamefont {Zhu}\ \emph {et~al.}(2021)\citenamefont {Zhu},
  \citenamefont {Shao}, \citenamefont {Yu}, \citenamefont {Cheng},
  \citenamefont {Desiatov}, \citenamefont {Xin}, \citenamefont {Hu},
  \citenamefont {Holzgrafe}, \citenamefont {Ghosh}, \citenamefont
  {Shams-Ansari}, \citenamefont {Puma}, \citenamefont {Sinclair}, \citenamefont
  {Reimer}, \citenamefont {Zhang},\ and\ \citenamefont {Lon\v{c}ar}}]{Zhu:21}%
  \BibitemOpen
  \bibfield  {author} {\bibinfo {author} {\bibfnamefont {D.}~\bibnamefont
  {Zhu}}, \bibinfo {author} {\bibfnamefont {L.}~\bibnamefont {Shao}}, \bibinfo
  {author} {\bibfnamefont {M.}~\bibnamefont {Yu}}, \bibinfo {author}
  {\bibfnamefont {R.}~\bibnamefont {Cheng}}, \bibinfo {author} {\bibfnamefont
  {B.}~\bibnamefont {Desiatov}}, \bibinfo {author} {\bibfnamefont {C.~J.}\
  \bibnamefont {Xin}}, \bibinfo {author} {\bibfnamefont {Y.}~\bibnamefont
  {Hu}}, \bibinfo {author} {\bibfnamefont {J.}~\bibnamefont {Holzgrafe}},
  \bibinfo {author} {\bibfnamefont {S.}~\bibnamefont {Ghosh}}, \bibinfo
  {author} {\bibfnamefont {A.}~\bibnamefont {Shams-Ansari}}, \bibinfo {author}
  {\bibfnamefont {E.}~\bibnamefont {Puma}}, \bibinfo {author} {\bibfnamefont
  {N.}~\bibnamefont {Sinclair}}, \bibinfo {author} {\bibfnamefont
  {C.}~\bibnamefont {Reimer}}, \bibinfo {author} {\bibfnamefont
  {M.}~\bibnamefont {Zhang}},\ and\ \bibinfo {author} {\bibfnamefont
  {M.}~\bibnamefont {Lon\v{c}ar}},\ }\bibfield  {title} {\bibinfo {title}
  {Integrated photonics on thin-film lithium niobate},\ }\href
  {https://doi.org/10.1364/AOP.411024} {\bibfield  {journal} {\bibinfo
  {journal} {Adv. Opt. Photon.}\ }\textbf {\bibinfo {volume} {13}},\ \bibinfo
  {pages} {242} (\bibinfo {year} {2021})}\BibitemShut {NoStop}%
\bibitem [{\citenamefont {Ledbetter}\ \emph {et~al.}(2004)\citenamefont
  {Ledbetter}, \citenamefont {Ogi},\ and\ \citenamefont
  {Nakamura}}]{LEDBETTER2004941}%
  \BibitemOpen
  \bibfield  {author} {\bibinfo {author} {\bibfnamefont {H.}~\bibnamefont
  {Ledbetter}}, \bibinfo {author} {\bibfnamefont {H.}~\bibnamefont {Ogi}},\
  and\ \bibinfo {author} {\bibfnamefont {N.}~\bibnamefont {Nakamura}},\
  }\bibfield  {title} {\bibinfo {title} {Elastic, anelastic, piezoelectric
  coefficients of monocrystal lithium niobate},\ }\href
  {https://doi.org/https://doi.org/10.1016/j.mechmat.2003.08.013} {\bibfield
  {journal} {\bibinfo  {journal} {Mechanics of Materials}\ }\textbf {\bibinfo
  {volume} {36}},\ \bibinfo {pages} {941} (\bibinfo {year} {2004})},\ \bibinfo
  {note} {active Materials}\BibitemShut {NoStop}%
\bibitem [{\citenamefont {Bajak}\ \emph {et~al.}(1981)\citenamefont {Bajak},
  \citenamefont {McNab}, \citenamefont {Richter},\ and\ \citenamefont
  {Wilkinson}}]{doi:10.1121/1.385588}%
  \BibitemOpen
  \bibfield  {author} {\bibinfo {author} {\bibfnamefont {I.~L.}\ \bibnamefont
  {Bajak}}, \bibinfo {author} {\bibfnamefont {A.}~\bibnamefont {McNab}},
  \bibinfo {author} {\bibfnamefont {J.}~\bibnamefont {Richter}},\ and\ \bibinfo
  {author} {\bibfnamefont {C.~D.~W.}\ \bibnamefont {Wilkinson}},\ }\bibfield
  {title} {\bibinfo {title} {Attenuation of acoustic waves in lithium
  niobate},\ }\href {https://doi.org/10.1121/1.385588} {\bibfield  {journal}
  {\bibinfo  {journal} {The Journal of the Acoustical Society of America}\
  }\textbf {\bibinfo {volume} {69}},\ \bibinfo {pages} {689} (\bibinfo {year}
  {1981})}\BibitemShut {NoStop}%
\bibitem [{\citenamefont {Desiatov}\ \emph {et~al.}(2019)\citenamefont
  {Desiatov}, \citenamefont {Shams-Ansari}, \citenamefont {Zhang},
  \citenamefont {Wang},\ and\ \citenamefont {Lon\v{c}ar}}]{Desiatov:19}%
  \BibitemOpen
  \bibfield  {author} {\bibinfo {author} {\bibfnamefont {B.}~\bibnamefont
  {Desiatov}}, \bibinfo {author} {\bibfnamefont {A.}~\bibnamefont
  {Shams-Ansari}}, \bibinfo {author} {\bibfnamefont {M.}~\bibnamefont {Zhang}},
  \bibinfo {author} {\bibfnamefont {C.}~\bibnamefont {Wang}},\ and\ \bibinfo
  {author} {\bibfnamefont {M.}~\bibnamefont {Lon\v{c}ar}},\ }\bibfield  {title}
  {\bibinfo {title} {Ultra-low-loss integrated visible photonics using
  thin-film lithium niobate},\ }\href {https://doi.org/10.1364/OPTICA.6.000380}
  {\bibfield  {journal} {\bibinfo  {journal} {Optica}\ }\textbf {\bibinfo
  {volume} {6}},\ \bibinfo {pages} {380} (\bibinfo {year} {2019})}\BibitemShut
  {NoStop}%
\bibitem [{\citenamefont {Wang}\ \emph {et~al.}(2017)\citenamefont {Wang},
  \citenamefont {Xiong}, \citenamefont {Andrade}, \citenamefont {Venkataraman},
  \citenamefont {Ren}, \citenamefont {Guo},\ and\ \citenamefont
  {Lon\v{c}ar}}]{Wang:17}%
  \BibitemOpen
  \bibfield  {author} {\bibinfo {author} {\bibfnamefont {C.}~\bibnamefont
  {Wang}}, \bibinfo {author} {\bibfnamefont {X.}~\bibnamefont {Xiong}},
  \bibinfo {author} {\bibfnamefont {N.}~\bibnamefont {Andrade}}, \bibinfo
  {author} {\bibfnamefont {V.}~\bibnamefont {Venkataraman}}, \bibinfo {author}
  {\bibfnamefont {X.-F.}\ \bibnamefont {Ren}}, \bibinfo {author} {\bibfnamefont
  {G.-C.}\ \bibnamefont {Guo}},\ and\ \bibinfo {author} {\bibfnamefont
  {M.}~\bibnamefont {Lon\v{c}ar}},\ }\bibfield  {title} {\bibinfo {title}
  {Second harmonic generation in nano-structured thin-film lithium niobate
  waveguides},\ }\href {https://doi.org/10.1364/OE.25.006963} {\bibfield
  {journal} {\bibinfo  {journal} {Opt. Express}\ }\textbf {\bibinfo {volume}
  {25}},\ \bibinfo {pages} {6963} (\bibinfo {year} {2017})}\BibitemShut
  {NoStop}%
\bibitem [{\citenamefont {Li}\ \emph {et~al.}(2022)\citenamefont {Li},
  \citenamefont {Wang}, \citenamefont {Lihachev}, \citenamefont {Tan},
  \citenamefont {Snigirev}, \citenamefont {Churaev}, \citenamefont {Kuznetsov},
  \citenamefont {Siddharth}, \citenamefont {Bereyhi}, \citenamefont
  {Riemensberger},\ and\ \citenamefont
  {Kippenberg}}]{https://doi.org/10.48550/arxiv.2208.05556}%
  \BibitemOpen
  \bibfield  {author} {\bibinfo {author} {\bibfnamefont {Z.}~\bibnamefont
  {Li}}, \bibinfo {author} {\bibfnamefont {R.~N.}\ \bibnamefont {Wang}},
  \bibinfo {author} {\bibfnamefont {G.}~\bibnamefont {Lihachev}}, \bibinfo
  {author} {\bibfnamefont {Z.}~\bibnamefont {Tan}}, \bibinfo {author}
  {\bibfnamefont {V.}~\bibnamefont {Snigirev}}, \bibinfo {author}
  {\bibfnamefont {M.}~\bibnamefont {Churaev}}, \bibinfo {author} {\bibfnamefont
  {N.}~\bibnamefont {Kuznetsov}}, \bibinfo {author} {\bibfnamefont
  {A.}~\bibnamefont {Siddharth}}, \bibinfo {author} {\bibfnamefont {M.~J.}\
  \bibnamefont {Bereyhi}}, \bibinfo {author} {\bibfnamefont {J.}~\bibnamefont
  {Riemensberger}},\ and\ \bibinfo {author} {\bibfnamefont {T.~J.}\
  \bibnamefont {Kippenberg}},\ }\href
  {https://doi.org/10.48550/ARXIV.2208.05556} {\bibinfo {title} {Tightly
  confining lithium niobate photonic integrated circuits and lasers}} (\bibinfo
  {year} {2022})\BibitemShut {NoStop}%
\bibitem [{\citenamefont {Zurita}\ \emph {et~al.}(2021)\citenamefont {Zurita},
  \citenamefont {Wiederhecker},\ and\ \citenamefont {Alegre}}]{Zurita:21}%
  \BibitemOpen
  \bibfield  {author} {\bibinfo {author} {\bibfnamefont {R.~O.}\ \bibnamefont
  {Zurita}}, \bibinfo {author} {\bibfnamefont {G.~S.}\ \bibnamefont
  {Wiederhecker}},\ and\ \bibinfo {author} {\bibfnamefont {T.~P.~M.}\
  \bibnamefont {Alegre}},\ }\bibfield  {title} {\bibinfo {title} {Designing of
  strongly confined short-wave brillouin phonons in silicon waveguide periodic
  lattices},\ }\href {https://doi.org/10.1364/OE.413770} {\bibfield  {journal}
  {\bibinfo  {journal} {Opt. Express}\ }\textbf {\bibinfo {volume} {29}},\
  \bibinfo {pages} {1736} (\bibinfo {year} {2021})}\BibitemShut {NoStop}%
\bibitem [{\citenamefont {Wang}\ \emph {et~al.}(2018)\citenamefont {Wang},
  \citenamefont {Zhang}, \citenamefont {Chen}, \citenamefont {Bertrand},
  \citenamefont {Shams-Ansari}, \citenamefont {Chandrasekhar}, \citenamefont
  {Winzer},\ and\ \citenamefont {Lon{\v{c}}ar}}]{Wang2018}%
  \BibitemOpen
  \bibfield  {author} {\bibinfo {author} {\bibfnamefont {C.}~\bibnamefont
  {Wang}}, \bibinfo {author} {\bibfnamefont {M.}~\bibnamefont {Zhang}},
  \bibinfo {author} {\bibfnamefont {X.}~\bibnamefont {Chen}}, \bibinfo {author}
  {\bibfnamefont {M.}~\bibnamefont {Bertrand}}, \bibinfo {author}
  {\bibfnamefont {A.}~\bibnamefont {Shams-Ansari}}, \bibinfo {author}
  {\bibfnamefont {S.}~\bibnamefont {Chandrasekhar}}, \bibinfo {author}
  {\bibfnamefont {P.}~\bibnamefont {Winzer}},\ and\ \bibinfo {author}
  {\bibfnamefont {M.}~\bibnamefont {Lon{\v{c}}ar}},\ }\bibfield  {title}
  {\bibinfo {title} {Integrated lithium niobate electro-optic modulators
  operating at cmos-compatible voltages},\ }\href
  {https://doi.org/10.1038/s41586-018-0551-y} {\bibfield  {journal} {\bibinfo
  {journal} {Nature}\ }\textbf {\bibinfo {volume} {562}},\ \bibinfo {pages}
  {101} (\bibinfo {year} {2018})}\BibitemShut {NoStop}%
\bibitem [{\citenamefont {Cortes-Herrera}\ \emph {et~al.}(2021)\citenamefont
  {Cortes-Herrera}, \citenamefont {He}, \citenamefont {Cardenas},\ and\
  \citenamefont {Agrawal}}]{Cortes-Herrera:21}%
  \BibitemOpen
  \bibfield  {author} {\bibinfo {author} {\bibfnamefont {L.}~\bibnamefont
  {Cortes-Herrera}}, \bibinfo {author} {\bibfnamefont {X.}~\bibnamefont {He}},
  \bibinfo {author} {\bibfnamefont {J.}~\bibnamefont {Cardenas}},\ and\
  \bibinfo {author} {\bibfnamefont {G.~P.}\ \bibnamefont {Agrawal}},\
  }\bibfield  {title} {\bibinfo {title} {Design of an x-cut thin-film lithium
  niobate waveguide as a passive polarization rotator},\ }\href
  {https://doi.org/10.1364/OE.445412} {\bibfield  {journal} {\bibinfo
  {journal} {Opt. Express}\ }\textbf {\bibinfo {volume} {29}},\ \bibinfo
  {pages} {44174} (\bibinfo {year} {2021})}\BibitemShut {NoStop}%
\bibitem [{\citenamefont {Hu}\ \emph {et~al.}(2021)\citenamefont {Hu},
  \citenamefont {Yu}, \citenamefont {Zhu}, \citenamefont {Sinclair},
  \citenamefont {Shams-Ansari}, \citenamefont {Shao}, \citenamefont
  {Holzgrafe}, \citenamefont {Puma}, \citenamefont {Zhang},\ and\ \citenamefont
  {Lon{\v{c}}ar}}]{Hu2021}%
  \BibitemOpen
  \bibfield  {author} {\bibinfo {author} {\bibfnamefont {Y.}~\bibnamefont
  {Hu}}, \bibinfo {author} {\bibfnamefont {M.}~\bibnamefont {Yu}}, \bibinfo
  {author} {\bibfnamefont {D.}~\bibnamefont {Zhu}}, \bibinfo {author}
  {\bibfnamefont {N.}~\bibnamefont {Sinclair}}, \bibinfo {author}
  {\bibfnamefont {A.}~\bibnamefont {Shams-Ansari}}, \bibinfo {author}
  {\bibfnamefont {L.}~\bibnamefont {Shao}}, \bibinfo {author} {\bibfnamefont
  {J.}~\bibnamefont {Holzgrafe}}, \bibinfo {author} {\bibfnamefont
  {E.}~\bibnamefont {Puma}}, \bibinfo {author} {\bibfnamefont {M.}~\bibnamefont
  {Zhang}},\ and\ \bibinfo {author} {\bibfnamefont {M.}~\bibnamefont
  {Lon{\v{c}}ar}},\ }\bibfield  {title} {\bibinfo {title} {On-chip
  electro-optic frequency shifters and beam splitters},\ }\href
  {https://doi.org/10.1038/s41586-021-03999-x} {\bibfield  {journal} {\bibinfo
  {journal} {Nature}\ }\textbf {\bibinfo {volume} {599}},\ \bibinfo {pages}
  {587} (\bibinfo {year} {2021})}\BibitemShut {NoStop}%
\bibitem [{\citenamefont {Shen}\ \emph {et~al.}(2020)\citenamefont {Shen},
  \citenamefont {Xie}, \citenamefont {Zou}, \citenamefont {Xu}, \citenamefont
  {Fu},\ and\ \citenamefont {Tang}}]{doi:10.1063/5.0020019}%
  \BibitemOpen
  \bibfield  {author} {\bibinfo {author} {\bibfnamefont {M.}~\bibnamefont
  {Shen}}, \bibinfo {author} {\bibfnamefont {J.}~\bibnamefont {Xie}}, \bibinfo
  {author} {\bibfnamefont {C.-L.}\ \bibnamefont {Zou}}, \bibinfo {author}
  {\bibfnamefont {Y.}~\bibnamefont {Xu}}, \bibinfo {author} {\bibfnamefont
  {W.}~\bibnamefont {Fu}},\ and\ \bibinfo {author} {\bibfnamefont {H.~X.}\
  \bibnamefont {Tang}},\ }\bibfield  {title} {\bibinfo {title} {High frequency
  lithium niobate film-thickness-mode optomechanical resonator},\ }\href
  {https://doi.org/10.1063/5.0020019} {\bibfield  {journal} {\bibinfo
  {journal} {Applied Physics Letters}\ }\textbf {\bibinfo {volume} {117}},\
  \bibinfo {pages} {131104} (\bibinfo {year} {2020})}\BibitemShut {NoStop}%
\bibitem [{\citenamefont {Marpaung}\ \emph {et~al.}(2015)\citenamefont
  {Marpaung}, \citenamefont {Morrison}, \citenamefont {Pagani}, \citenamefont
  {Pant}, \citenamefont {Choi}, \citenamefont {Luther-Davies}, \citenamefont
  {Madden},\ and\ \citenamefont {Eggleton}}]{Marpaung:15}%
  \BibitemOpen
  \bibfield  {author} {\bibinfo {author} {\bibfnamefont {D.}~\bibnamefont
  {Marpaung}}, \bibinfo {author} {\bibfnamefont {B.}~\bibnamefont {Morrison}},
  \bibinfo {author} {\bibfnamefont {M.}~\bibnamefont {Pagani}}, \bibinfo
  {author} {\bibfnamefont {R.}~\bibnamefont {Pant}}, \bibinfo {author}
  {\bibfnamefont {D.-Y.}\ \bibnamefont {Choi}}, \bibinfo {author}
  {\bibfnamefont {B.}~\bibnamefont {Luther-Davies}}, \bibinfo {author}
  {\bibfnamefont {S.~J.}\ \bibnamefont {Madden}},\ and\ \bibinfo {author}
  {\bibfnamefont {B.~J.}\ \bibnamefont {Eggleton}},\ }\bibfield  {title}
  {\bibinfo {title} {Low-power, chip-based stimulated brillouin scattering
  microwave photonic filter with ultrahigh selectivity},\ }\href
  {https://doi.org/10.1364/OPTICA.2.000076} {\bibfield  {journal} {\bibinfo
  {journal} {Optica}\ }\textbf {\bibinfo {volume} {2}},\ \bibinfo {pages} {76}
  (\bibinfo {year} {2015})}\BibitemShut {NoStop}%
\bibitem [{\citenamefont {Spirin}\ \emph {et~al.}(2006)\citenamefont {Spirin},
  \citenamefont {Kellerman}, \citenamefont {Swart},\ and\ \citenamefont
  {Fotiadi}}]{Spirin:06}%
  \BibitemOpen
  \bibfield  {author} {\bibinfo {author} {\bibfnamefont {V.~V.}\ \bibnamefont
  {Spirin}}, \bibinfo {author} {\bibfnamefont {J.}~\bibnamefont {Kellerman}},
  \bibinfo {author} {\bibfnamefont {P.~L.}\ \bibnamefont {Swart}},\ and\
  \bibinfo {author} {\bibfnamefont {A.~A.}\ \bibnamefont {Fotiadi}},\
  }\bibfield  {title} {\bibinfo {title} {Intensity noise in sbs with injection
  locking generation of stokes seed signal},\ }\href
  {https://doi.org/10.1364/OE.14.008328} {\bibfield  {journal} {\bibinfo
  {journal} {Opt. Express}\ }\textbf {\bibinfo {volume} {14}},\ \bibinfo
  {pages} {8328} (\bibinfo {year} {2006})}\BibitemShut {NoStop}%
\bibitem [{\citenamefont {Zelmon}\ \emph {et~al.}(1997)\citenamefont {Zelmon},
  \citenamefont {Small},\ and\ \citenamefont {Jundt}}]{Zelmon:97}%
  \BibitemOpen
  \bibfield  {author} {\bibinfo {author} {\bibfnamefont {D.~E.}\ \bibnamefont
  {Zelmon}}, \bibinfo {author} {\bibfnamefont {D.~L.}\ \bibnamefont {Small}},\
  and\ \bibinfo {author} {\bibfnamefont {D.}~\bibnamefont {Jundt}},\ }\bibfield
   {title} {\bibinfo {title} {Infrared corrected sellmeier coefficients for
  congruently grown lithium niobate and 5 mol.$\%$ magnesium oxide--doped
  lithium niobate},\ }\href {https://doi.org/10.1364/JOSAB.14.003319}
  {\bibfield  {journal} {\bibinfo  {journal} {J. Opt. Soc. Am. B}\ }\textbf
  {\bibinfo {volume} {14}},\ \bibinfo {pages} {3319} (\bibinfo {year}
  {1997})}\BibitemShut {NoStop}%
\bibitem [{\citenamefont {Auld}(1973)}]{auld1973acoustic}%
  \BibitemOpen
  \bibfield  {author} {\bibinfo {author} {\bibfnamefont {B.}~\bibnamefont
  {Auld}},\ }\href {https://books.google.com.br/books?id=\_2MWAwAAQBAJ} {\emph
  {\bibinfo {title} {Acoustic fields and waves in solids}}},\ A
  Wiley-Interscience publication\ (\bibinfo  {publisher} {Wiley},\ \bibinfo
  {year} {1973})\BibitemShut {NoStop}%
\bibitem [{\citenamefont {Nelson}\ and\ \citenamefont
  {Lindsay}(1984)}]{doi:10.1121/1.390547}%
  \BibitemOpen
  \bibfield  {author} {\bibinfo {author} {\bibfnamefont {D.~F.}\ \bibnamefont
  {Nelson}}\ and\ \bibinfo {author} {\bibfnamefont {R.~B.}\ \bibnamefont
  {Lindsay}},\ }\bibfield  {title} {\bibinfo {title} {Electric, optic, and
  acoustic interactions in dielectrics by donald f. nelson},\ }\href
  {https://doi.org/10.1121/1.390547} {\bibfield  {journal} {\bibinfo  {journal}
  {The Journal of the Acoustical Society of America}\ }\textbf {\bibinfo
  {volume} {75}},\ \bibinfo {pages} {646} (\bibinfo {year} {1984})}\BibitemShut
  {NoStop}%
\bibitem [{\citenamefont {Mytsyk}\ \emph {et~al.}(2021)\citenamefont {Mytsyk},
  \citenamefont {Demyanyshyn}, \citenamefont {Andrushchak},\ and\ \citenamefont
  {Buryy}}]{cryst11091095}%
  \BibitemOpen
  \bibfield  {author} {\bibinfo {author} {\bibfnamefont {B.}~\bibnamefont
  {Mytsyk}}, \bibinfo {author} {\bibfnamefont {N.}~\bibnamefont {Demyanyshyn}},
  \bibinfo {author} {\bibfnamefont {A.}~\bibnamefont {Andrushchak}},\ and\
  \bibinfo {author} {\bibfnamefont {O.}~\bibnamefont {Buryy}},\ }\bibfield
  {title} {\bibinfo {title} {Photoelastic properties of trigonal crystals},\
  }\bibfield  {journal} {\bibinfo  {journal} {Crystals}\ }\textbf {\bibinfo
  {volume} {11}},\ \href {https://doi.org/10.3390/cryst11091095}
  {10.3390/cryst11091095} (\bibinfo {year} {2021})\BibitemShut {NoStop}%
\bibitem [{\citenamefont
  {Narasimhamurty}(2012)}]{narasimhamurty2012photoelastic}%
  \BibitemOpen
  \bibfield  {author} {\bibinfo {author} {\bibfnamefont {T.}~\bibnamefont
  {Narasimhamurty}},\ }\href
  {https://books.google.com.br/books?id=iCsLCAAAQBAJ} {\emph {\bibinfo {title}
  {Photoelastic and Electro-Optic Properties of Crystals}}}\ (\bibinfo
  {publisher} {Springer US},\ \bibinfo {year} {2012})\BibitemShut {NoStop}%
\bibitem [{\citenamefont {Kottke}\ \emph {et~al.}(2008)\citenamefont {Kottke},
  \citenamefont {Farjadpour},\ and\ \citenamefont
  {Johnson}}]{PhysRevE.77.036611}%
  \BibitemOpen
  \bibfield  {author} {\bibinfo {author} {\bibfnamefont {C.}~\bibnamefont
  {Kottke}}, \bibinfo {author} {\bibfnamefont {A.}~\bibnamefont {Farjadpour}},\
  and\ \bibinfo {author} {\bibfnamefont {S.~G.}\ \bibnamefont {Johnson}},\
  }\bibfield  {title} {\bibinfo {title} {Perturbation theory for anisotropic
  dielectric interfaces, and application to subpixel smoothing of discretized
  numerical methods},\ }\href {https://doi.org/10.1103/PhysRevE.77.036611}
  {\bibfield  {journal} {\bibinfo  {journal} {Phys. Rev. E}\ }\textbf {\bibinfo
  {volume} {77}},\ \bibinfo {pages} {036611} (\bibinfo {year}
  {2008})}\BibitemShut {NoStop}%
\bibitem [{\citenamefont {Abedin}(2005)}]{S.Abedin:05}%
  \BibitemOpen
  \bibfield  {author} {\bibinfo {author} {\bibfnamefont {K.~S.}\ \bibnamefont
  {Abedin}},\ }\bibfield  {title} {\bibinfo {title} {Observation of strong
  stimulated brillouin scattering in single-mode as2se3 chalcogenide fiber},\
  }\href {https://doi.org/10.1364/OPEX.13.010266} {\bibfield  {journal}
  {\bibinfo  {journal} {Opt. Express}\ }\textbf {\bibinfo {volume} {13}},\
  \bibinfo {pages} {10266} (\bibinfo {year} {2005})}\BibitemShut {NoStop}%
\bibitem [{\citenamefont {Choudhary}\ \emph {et~al.}(2017)\citenamefont
  {Choudhary}, \citenamefont {Morrison}, \citenamefont {Aryanfar},
  \citenamefont {Shahnia}, \citenamefont {Pagani}, \citenamefont {Liu},
  \citenamefont {Vu}, \citenamefont {Madden}, \citenamefont {Marpaung},\ and\
  \citenamefont {Eggleton}}]{Choudhary:17}%
  \BibitemOpen
  \bibfield  {author} {\bibinfo {author} {\bibfnamefont {A.}~\bibnamefont
  {Choudhary}}, \bibinfo {author} {\bibfnamefont {B.}~\bibnamefont {Morrison}},
  \bibinfo {author} {\bibfnamefont {I.}~\bibnamefont {Aryanfar}}, \bibinfo
  {author} {\bibfnamefont {S.}~\bibnamefont {Shahnia}}, \bibinfo {author}
  {\bibfnamefont {M.}~\bibnamefont {Pagani}}, \bibinfo {author} {\bibfnamefont
  {Y.}~\bibnamefont {Liu}}, \bibinfo {author} {\bibfnamefont {K.}~\bibnamefont
  {Vu}}, \bibinfo {author} {\bibfnamefont {S.}~\bibnamefont {Madden}}, \bibinfo
  {author} {\bibfnamefont {D.}~\bibnamefont {Marpaung}},\ and\ \bibinfo
  {author} {\bibfnamefont {B.~J.}\ \bibnamefont {Eggleton}},\ }\bibfield
  {title} {\bibinfo {title} {Advanced integrated microwave signal processing
  with giant on-chip brillouin gain},\ }\href
  {https://opg.optica.org/jlt/abstract.cfm?URI=jlt-35-4-846} {\bibfield
  {journal} {\bibinfo  {journal} {J. Lightwave Technol.}\ }\textbf {\bibinfo
  {volume} {35}},\ \bibinfo {pages} {846} (\bibinfo {year} {2017})}\BibitemShut
  {NoStop}%
\bibitem [{\citenamefont {Gundavarapu}\ \emph {et~al.}(2019)\citenamefont
  {Gundavarapu}, \citenamefont {Brodnik}, \citenamefont {Puckett},
  \citenamefont {Huffman}, \citenamefont {Bose}, \citenamefont {Behunin},
  \citenamefont {Wu}, \citenamefont {Qiu}, \citenamefont {Pinho}, \citenamefont
  {Chauhan}, \citenamefont {Nohava}, \citenamefont {Rakich}, \citenamefont
  {Nelson}, \citenamefont {Salit},\ and\ \citenamefont
  {Blumenthal}}]{Gundavarapu2019}%
  \BibitemOpen
  \bibfield  {author} {\bibinfo {author} {\bibfnamefont {S.}~\bibnamefont
  {Gundavarapu}}, \bibinfo {author} {\bibfnamefont {G.~M.}\ \bibnamefont
  {Brodnik}}, \bibinfo {author} {\bibfnamefont {M.}~\bibnamefont {Puckett}},
  \bibinfo {author} {\bibfnamefont {T.}~\bibnamefont {Huffman}}, \bibinfo
  {author} {\bibfnamefont {D.}~\bibnamefont {Bose}}, \bibinfo {author}
  {\bibfnamefont {R.}~\bibnamefont {Behunin}}, \bibinfo {author} {\bibfnamefont
  {J.}~\bibnamefont {Wu}}, \bibinfo {author} {\bibfnamefont {T.}~\bibnamefont
  {Qiu}}, \bibinfo {author} {\bibfnamefont {C.}~\bibnamefont {Pinho}}, \bibinfo
  {author} {\bibfnamefont {N.}~\bibnamefont {Chauhan}}, \bibinfo {author}
  {\bibfnamefont {J.}~\bibnamefont {Nohava}}, \bibinfo {author} {\bibfnamefont
  {P.~T.}\ \bibnamefont {Rakich}}, \bibinfo {author} {\bibfnamefont {K.~D.}\
  \bibnamefont {Nelson}}, \bibinfo {author} {\bibfnamefont {M.}~\bibnamefont
  {Salit}},\ and\ \bibinfo {author} {\bibfnamefont {D.~J.}\ \bibnamefont
  {Blumenthal}},\ }\bibfield  {title} {\bibinfo {title} {Sub-hertz fundamental
  linewidth photonic integrated brillouin laser},\ }\href
  {https://doi.org/10.1038/s41566-018-0313-2} {\bibfield  {journal} {\bibinfo
  {journal} {Nature Photonics}\ }\textbf {\bibinfo {volume} {13}},\ \bibinfo
  {pages} {60} (\bibinfo {year} {2019})}\BibitemShut {NoStop}%
\end{thebibliography}%

\section{Supplementary Information\label{supp}}

The parameters used to estimate $G_{B}/Q_{m}$ in \Cref{sec:eng_photo} are shown in \Cref{tab:LN_typical_params}, and a full comparison between common Brillouin materials is shown in \Cref{tab:GB_literature_values} at end of the document.

\begin{table}[ht]
    \centering
    \caption{Parameters used to estimate $G_{B}/Q_{m}$ at \Cref{sec:eng_photo} for the LN system.\vspace{0.2cm}}
    \label{tab:LN_typical_params}
    \begin{tabular}{clcc}
        \toprule
        Symbol & Description & Value & Unit \\
        \midrule
        $c$ & Speed of light & 299792458 & m/s\\
        \addlinespace
        $\omega_{p}/2\pi$ & Optical frequency & 193 & THz\\
        \addlinespace
        $\rho$ & Material's density & 4643 & kg/m$^{3}$ \\
        \addlinespace
        $\Omega_{m}/2\pi$ & Mechanical frequency & 7 & GHz \\
        \addlinespace
        $A_{\text{eff}}$ & Effective optical area & 0.4 & $\mu $m$^{2}$\\
        \addlinespace
        $n_{\text{eff}}$ & Effective refractive index & 1.7 & 1 \\
        \addlinespace
        $\beta_{m}$ & Acoustic wavevector & 14 & rad/$\mu$m \\
        \bottomrule
    \end{tabular}
\end{table}

\subsection{Mechanical polarization colors}\label{sec:color_mechpol}

The acoustic displacement field $\vec{u} = (u_{x},u_{y},u_{z})$ is used to define the RGB color spectrum shown at \Cref{fig:LN_Z-cut_Dispersions_and_GB} and \Cref{fig:LN_X-cut_Dispersions_and_GB} on the mechanical dispersion. The RGB (red, green and blue) field $\hat{u}_{\text{RGB}}$ which paints the mechanical polarization is defined as

\begin{equation}
    \hat{u}_{\text{RGB}} = (u_{\text{R}},u_{\text{G}},u_{\text{B}}) := \frac{\int (u_{x}^{2},u_{y}^{2},u_{z}^{2}) dA}{\int (u_{x}^{2} + u_{y}^{2} + u_{z}^{2}) dA},
\end{equation}

\noindent which is basically a weighted average between the displacement directions. We then obtain three functions which sum to one, allowing us to construct our own color palette after creating a bijection to the set [0,255].

\subsection{Optical Properties}

The linear permittivity tensor $\varepsilon_{ij}$ is the relation between the electric displacement vector $D_{i}$ and the electric field $E_{j}$ in the frequency domain, $D_{i}(\omega) = \varepsilon_{ij}(\omega)E_{j}(\omega)$, where we can write $\varepsilon_{ij}(\omega)$ using as basis the principal axes of the crystal as

\begin{equation}
    \left[\varepsilon_{ij}\right] =
    \begin{pmatrix}
        \varepsilon_{11} & 0 & 0\\
        0 & \varepsilon_{11} & 0\\
        0 & 0 & \varepsilon_{33}
    \end{pmatrix}
    = \varepsilon_{0}
    \begin{pmatrix}
        n_{o}^{2} & 0 & 0\\
        0 & n_{o}^{2} & 0\\
        0 & 0 & n_{e}^{2}
    \end{pmatrix},
\end{equation}

\noindent with $n_{o} = 2.21$ and $n_{e} = 2.13$ at $\lambda = 1550$ nm for LN~\cite{Zelmon:97}. However, the coupling between mechanical and electrical properties forces us to define two different permittivities $\varepsilon_{ij}^{S}$ and $\varepsilon_{ij}^{T}$, where the upper indexes $S$ and $T$ denotes constants strain and constant stress, respectively. 
This is not so important at optical frequencies but at RF frequencies their values can change by a factor more than two~\cite{doi:10.1063/1.1660528, Jazbinsek2002, Weis1985}.

\subsection{Mechanical Properties}

The dynamic of an anisotropic crystal is given by the elastodynamic equation (known as Navier-Cauchy equation in the case of isotropic materials) as

\begin{equation}
    \rho \partial_{t}^{2}u_{i} = \partial_{j}T_{ij},
\end{equation}

\noindent where $u_{i}$ is the mechanical displacement of the body, $\rho$ is the unperturbed density of the material ($\rho = 4643$ kg/m$^{3}$ for LN) and $T_{ij}$ is the stress tensor of the system. The stress tensor $T_{ij}$ is related to the strain tensor $S_{ij}$ through the constitutive relation

\begin{equation}
    T_{ij} = C_{ijkl}S_{kl} \quad \Rightarrow \quad T_{I} = C_{IJ}S_{J} 
\end{equation}

\noindent where $C_{ijkl}$ is the stiffness tensor. The capital indexes $I$ and $J$ are indexes in Voigt notation~\cite{auld1973acoustic, doi:10.1121/1.390547}. The tensor $C_{IJ}$ for LN~\cite{Weis1985}, in units of [GPa], is given (in the crystal's principal axes) by

\begin{equation*}
    \left[C_{IJ}\right] =
    \begin{pmatrix}
        c_{11} & c_{12} & c_{13} & c_{14} & 0 & 0\\
        c_{12} & c_{11} & c_{13} & -c_{14} & 0 & 0\\
        c_{13} & c_{13} & c_{33} & 0 & 0 & 0\\
        c_{14} & -c_{14} & 0 & c_{44} & 0 & 0\\
        0 & 0 & 0 & 0 & c_{44} & c_{14}\\
        0 & 0 & 0 & 0 & c_{14} & \dfrac{c_{11}-c_{12}}{2}
    \end{pmatrix}
    =
\end{equation*}

\begin{equation}
    =
    \begin{pmatrix}
        198.83 & 54.64 & 68.23 & 7.83 & 0 & 0\\
        54.64 & 198.83 & 68.23 & -7.83 & 0 & 0\\
        68.23 & 68.23 & 235.71 & 0 & 0 & 0\\
        7.83 & -7.83 & 0 & 59.86 & 0 & 0\\
        0 & 0 & 0 & 0 & 59.86 & 7.83\\
        0 & 0 & 0 & 0 & 7.83 & 72.095
    \end{pmatrix}.
\end{equation}

\subsection{Photoelastic Properties}\label{sec:photoelastic_props}

The photoelastic effect (also known as elastooptical effect) is represented by the photoelastic tensor $p_{ijkl}$~\cite{Weis1985, cryst11091095, narasimhamurty2012photoelastic} which in Voigt notation has the values (in the crystal's principal axes),

\begin{equation*}
    \left[p_{IJ}\right] =
    \begin{pmatrix}
        p_{11} & p_{12} & p_{13} & p_{14} & 0 & 0\\
        p_{12} & p_{11} & p_{13} & -p_{14} & 0 & 0\\
        p_{31} & p_{31} & p_{33} & 0 & 0 & 0\\
        p_{41} & -p_{41} & 0 & p_{44} & 0 & 0\\
        0 & 0 & 0 & 0 & p_{44} & p_{41}\\
        0 & 0 & 0 & 0 & p_{14} & \dfrac{p_{11}-p_{12}}{2}
    \end{pmatrix}
    =
\end{equation*}

\begin{equation}
    =
    \begin{pmatrix}
        -0.026 & 0.088 & 0.134 & -0.083 & 0 & 0\\
        0.088 & -0.026 & 0.134 & 0.083 & 0 & 0\\
        0.177 & 0.177 & 0.07 & 0 & 0 & 0\\
        -0.151 & 0.151 & 0 & 0.145 & 0 & 0\\
        0 & 0 & 0 & 0 & 0.145 & -0.151\\
        0 & 0 & 0 & 0 & -0.083 & -0.057
    \end{pmatrix}.
\end{equation}

Note that the photoelastic tensor does \textbf{not} have the same symmetry of the stiffness tensor, i.e.,  $[C_{IJ}]^{T} = [C_{JI}] = [C_{IJ}]$ but $[p_{IJ}]^{T} = [p_{JI}] \neq [p_{IJ}]$.

\subsection{Tensorial Rotations}

As we studied LN in many different crystallographic orientations, lots of rotations were made on all the material properties. Given a coordinate transformation $[a]$ which we will explicitly represent as

\begin{equation}
    [a] =
    \begin{pmatrix}
        a_{xx} & a_{xy} & a_{xz}\\
        a_{yx} & a_{yy} & a_{yz}\\
        a_{zx} & a_{zy} & a_{zz}
    \end{pmatrix},
\end{equation}

\noindent where $[a]^{-1} = [a]^{\text{T}}$ and $\det{\left(a\right)} = 1$. It is easy to rotate each of the tensors $\varepsilon_{ij}$, $C_{ijkl}$ and $p_{ijkl}$ into new tensors $\varepsilon_{ij}'$, $C_{ijkl}'$ and $p_{ijkl}'$, it is purely a linear algebra exercise. However, as it is practical to use Voigt notation most of the time, it is convenient to rotate the tensor in that representation directly. The details of the following steps can be checked in Chapter 3B, 3C and 3D of~\cite{auld1973acoustic}. The Voigt rotation matrix $[M]$ that we need is given by

\begin{widetext}
\begin{equation}
[M] =
\begin{pmatrix}
    &
    \begin{matrix}
        a_{xx}^{2} & a_{xy}^{2} & a_{xz}^{2}\\
        a_{yx}^{2} & a_{yy}^{2} & a_{yz}^{2}\\
        a_{zx}^{2} & a_{zy}^{2} & a_{zz}^{2}
    \end{matrix}
    & \vline &
    \begin{matrix}
        2a_{xy}a_{xz} & 2a_{xz}a_{xx} & 2a_{xx}a_{xy}\\
        2a_{yy}a_{yz} & 2a_{yz}a_{yx} & 2a_{yx}a_{yy}\\
        2a_{zy}a_{zz} & 2a_{zz}a_{zx} & 2a_{zx}a_{zy}
    \end{matrix}\\
    \hline &
    \begin{matrix}
        a_{yx}a_{zx} & a_{yy}a_{zy} & a_{yz}a_{zz}\\
        a_{zx}a_{xx} & a_{zy}a_{xy} & a_{zz}a_{xz}\\
        a_{xx}a_{yx} & a_{xy}a_{yy} & a_{xz}a_{yz}
    \end{matrix}
    & \vline &
    \begin{matrix}
        a_{yy}a_{zz} + a_{yz}a_{zy} & a_{yx}a_{zz} + a_{yz}a_{zx} & a_{yy}a_{zx} + a_{yx}a_{zy}\\
        a_{xy}a_{zz} + a_{xz}a_{zy} & a_{xz}a_{zx} + a_{xx}a_{zz} & a_{xx}a_{zy} + a_{xy}a_{zx}\\
        a_{xy}a_{yz} + a_{xz}a_{yy} & a_{xz}a_{yx} + a_{xx}a_{yz} & a_{xx}a_{yy} + a_{xy}a_{yx}
    \end{matrix}
\end{pmatrix}
\end{equation}
\end{widetext}

\noindent then those material properties can be rotate directly as

\begin{align}
    [\varepsilon_{ij}^{'}] &= [a][\varepsilon_{ij}][a]^{\text{T}}, \\ [C_{IJ}^{'}] &= [M][C_{IJ}][M]^{\text{T}}, \\ [p_{IJ}^{'}] &= [M][p_{IJ}][M]^{\text{T}},
\end{align}

\noindent where the rotation matrix $[a]$ is a rotation of the material properties around the fixed coordinate $y$-axis, i.e.,

\begin{equation}
    [a] =
    \begin{pmatrix}
        \cos{\theta} & 0 & \sin{\theta}\\
        0 & 1 & 0\\
        -\sin{\theta} & 0 & \cos{\theta}
    \end{pmatrix},
\end{equation}

\noindent where $\theta$ can be $\theta_{Z}$ or $\theta_{X}$ depending on the cut in analysis.

\subsection{Optical force densities}\label{optical_force_densities}

All the force densities have the same structure~\cite{doi:10.1063/1.5088169, Wolff:21} given by

\begin{align}
    &\label{force_density_PE} f_{\text{PE}} = \frac{1}{\sqrt{P_{p}P_{s}}\max{|\vec{u}|}}\sum_{i=1}^{3}\sum_{j=1}^{3}E_{i,p}^{*}\delta\varepsilon^{\text{PE}}_{ij}E_{j,s} \\ &\label{force_density_RO} f_{\text{RO}} = \frac{1}{\sqrt{P_{p}P_{s}}\max{|\vec{u}|}}\sum_{i=1}^{3}\sum_{j=1}^{3}E_{i,p}^{*}\delta\varepsilon^{\text{RO}}_{ij}E_{j,s}
\end{align}

\begin{figure*}[ht]
    \centering
    \includegraphics[width=16cm]{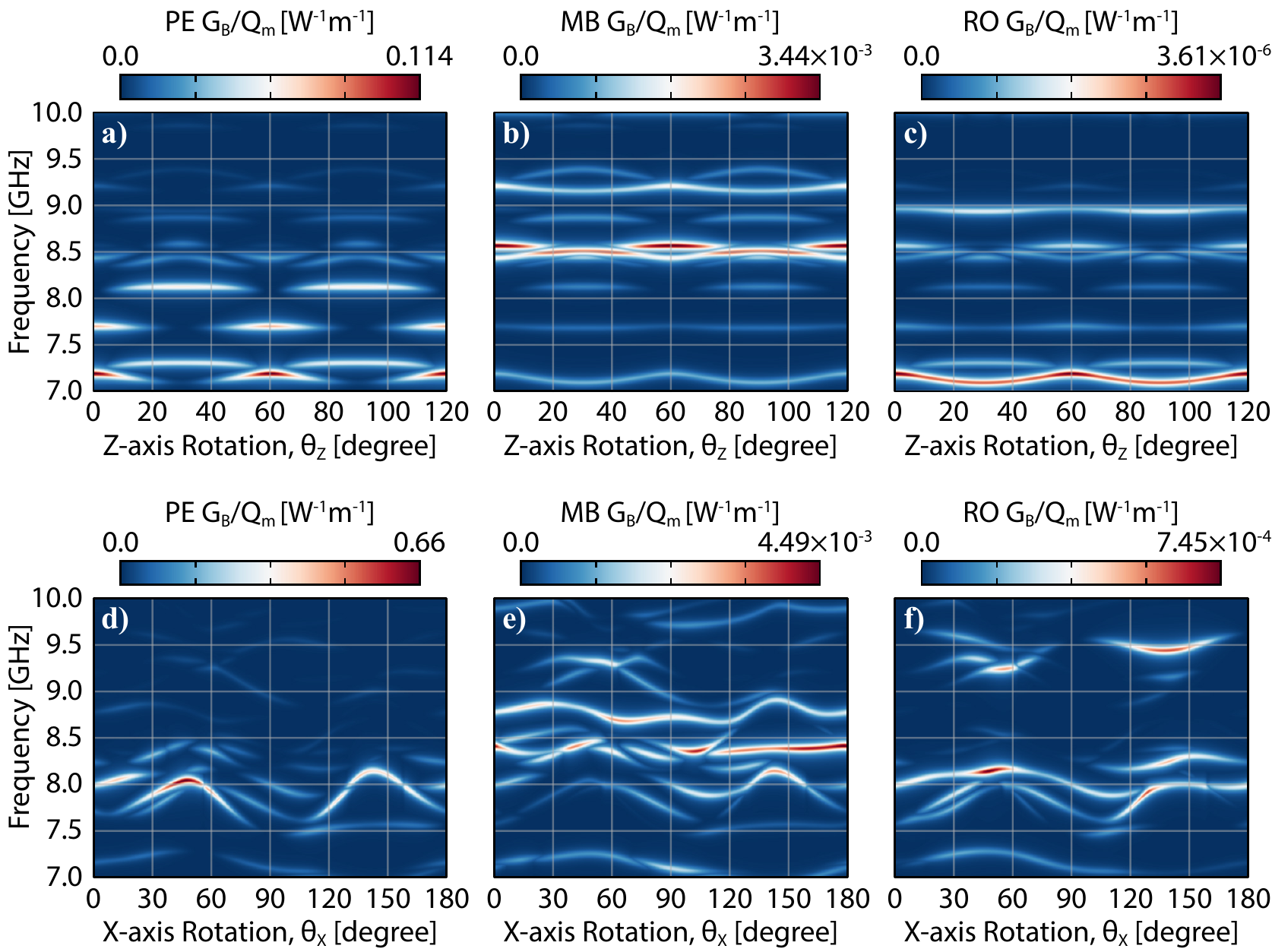}
    \caption{Separated contributions of the normalized Brillouin gain for the floating waveguide. \textbf{a)}-\textbf{b)}-\textbf{c)} $Z$-cut gain contributions shown in \Cref{fig:LN_Z-cut_Dispersions_and_GB}(d). From left to right: photoelastic, moving boundary and roto-optic; \textbf{d)}-\textbf{e)}-\textbf{f)} $X$-cut gain contributions shown in \Cref{fig:LN_X-cut_Dispersions_and_GB}(d).}
    \label{fig:LN_PE_MB_RO}
\end{figure*}

\noindent where the normalization term $P_{\gamma}$ ($\gamma \in \{p,s\}$), as well $m_{\text{eff}}$ mentioned in the beginning of the article, are given by

\begin{align}
    P_{\gamma} &= 2\Re\bigg\{\int\left(\vec{E}_{\gamma}\times\vec{H}_{\gamma}^{*}\right)\cdot\hat{z}dA\bigg\}, \\ m_{\text{eff}} &= \int \rho \left(\frac{|\vec{u}|}{\max|\vec{u}|}\right)^{2}dA.
\end{align}

The moving boundary contribution has a different treatment as it is an integral over the boundary but is completely explained at~\cite{PhysRevE.77.036611}, which is the article we have chosen to implement here. \Cref{force_density_PE} and \Cref{force_density_RO} tell us that it is just a matter of knowing the variation in the permittivity, $\delta\varepsilon_{ij}$, for each contribution to completely describe the force densities. The permittivity variation for each contribution~\cite{doi:10.1121/1.390547} is written as

\begin{align}
    \delta\varepsilon_{ij}^{\text{PE}} &= -\frac{1}{\varepsilon_{0}}\varepsilon_{ik}p_{klmn}S_{mn}\varepsilon_{lj}, \\ \delta\varepsilon_{ij}^{\text{RO}} &= R_{ik}\varepsilon_{kj} - \varepsilon_{ik}R_{kj}
\end{align}

\noindent where $S_{ij}$ and $R_{ij}$ are the elastic strain and the elastic rotation tensor, respectively, given by

\begin{align}
    S_{ij} &= \frac{1}{2}\left(\partial_{j}u_{i} + \partial_{i}u_{j}\right), \\ R_{ij} &= \frac{1}{2}\left(\partial_{j}u_{i} - \partial_{i}u_{j}\right).
\end{align}

While the photoelastic contribution directly depends on the photoelastic tensor $p_{klmn}$, the roto-optic contribution depends on the optical anisotropy of the system. In the case of optical isotropic material we always have $\delta\varepsilon_{ij}^{\text{RO}} = 0$ which can be easily proven just remembering that $\varepsilon_{ij} = \varepsilon_{0}n^2\delta_{ij}$, where $\varepsilon_{0}$ is the vacuum permittivity, $n$ is the refractive index and $\delta_{ij}$ is the Kronecker delta tensor. The contribution of each effect to the gain shown in \Cref{fig:LN_Z-cut_Dispersions_and_GB}(d) and \Cref{fig:LN_X-cut_Dispersions_and_GB}(d) can be seen at \Cref{supp} \Cref{fig:LN_PE_MB_RO}.

\addtolength{\tabcolsep}{3pt}
\begin{sidewaystable}[ht]
    \centering
    \vspace{8.5cm}
    \caption{Backward stimulated Brillouin scattering summary among standard materials and platforms.}
    \vspace{0.2cm}
    \label{tab:GB_literature_values}
    \begin{tabular}{cccccccc}
        \toprule
        Material & \begin{tabular}{c} Density\\ $[$kg/m$^{3}]$\end{tabular} & \begin{tabular}{c}Opt. Index\\ at 1550nm\end{tabular} & \begin{tabular}{c}Acoustic Phase\\Velocities $[$m/s$]$\end{tabular} & \begin{tabular}{c}Photoelastic\\ Coefficients\end{tabular} & \begin{tabular}{c}$\Omega_{m}/2\pi$\\ $[$GHz$]$\end{tabular} & $Q_{m}$ & \begin{tabular}{c} $G_{B}/Q_{m}$\\ $[$W$^{-1}$m$^{-1}]$\end{tabular} \\
        
        \midrule
        SiO$_{2}$ & 2203 & 1.44 & \begin{tabular}{l}$V_{S}=3763$\\ $V_{L}=5969$\end{tabular} & \begin{tabular}{l}$p_{11}=0.121$\\ $p_{12}=0.271$\end{tabular} & 12.81 & 357.93 & 0.00026~\cite{doi:10.1063/1.90726}\\

        \midrule
        As$_{2}$Se$_{3}$ & 4640 & 2.808 & \begin{tabular}{l}$V_{S}=1227$\\ $V_{L}=2250$\end{tabular} & \begin{tabular}{l}$p_{11}=0.314$\\ $p_{12}=0.266$\end{tabular} & 7.95 & 602.27 & 0.256~\cite{S.Abedin:05}\\

        \midrule
        As$_{2}$S$_{3}$ & 3210 & 2.437 & \begin{tabular}{l}$V_{S}=1401$ \\ $V_{L}=2555$\end{tabular} & \begin{tabular}{l}$p_{11}=0.25$\\ $p_{12}=0.24$\end{tabular} & \begin{tabular}{c}7.7\\ 7.6\end{tabular} & \begin{tabular}{c}226.47\\ 760\end{tabular} & \begin{tabular}{l}1.373~\cite{Pant:11}\\ 0.658~\cite{Choudhary:17}\end{tabular}\\

        \midrule
        Si & 2329 & 3.49 &
        \begin{tabular}{cc}
        $\left[100\right]$ & $\left[110\right]$\\
        \midrule
        5850 & 4680\\
        5850 & 5850\\
        8440 & 9130
        \end{tabular} & \begin{tabular}{l}$p_{11}=-0.094$\\ $p_{12}=0.017$\\ $p_{44}=-0.051$\end{tabular} & 13.66 & 970.86 & 0.37~\cite{VanLaer2015}\\

        \midrule
        Si$_{3}$N$_{4}$ & 3290 & 2.00 & \begin{tabular}{l}$V_{S}=6090$ \\ $V_{L}=10850$\end{tabular} & \begin{tabular}{l}$p_{11}=\text{Unknown}$\\ $|p_{12}|=0.047$\end{tabular} & \begin{tabular}{c}10.93\\ 25\\ 12.93\end{tabular} & \begin{tabular}{c}71.44\\ 64.10\\ 99.46\end{tabular} & \begin{tabular}{r}0.0014~\cite{Gundavarapu2019}\\ 0.0052~\cite{PhysRevLett.124.013902}\\ 0.0040~\cite{doi:10.1126/sciadv.abq2196}\end{tabular}\\

        \midrule
        LiNbO$_{3}$ & 4643 & \begin{tabular}{c}2.21 (o)\\ 2.13 (e)\end{tabular} &
        \begin{tabular}{ccc}
        $\left[100\right]$ & $\left[010\right]$ & $\left[001\right]$\\
        \midrule
        3474.2 & 3577.4 & 3590.6\\
        4043.5 & 3940.5 & 3590.6\\
        6544.0 & 6551.2 & 7125.1
        \end{tabular} &
        \begin{tabular}{ll} $p_{11}=-0.026$ & $p_{12}=0.088$\\ $p_{13}=0.134$ & $p_{14}=-0.083$\\ $p_{31}=0.177$ & $p_{33}=0.07$\\ $p_{41}=-0.151$ & $p_{44}=0.145$
        \end{tabular} & \begin{tabular}{c} 7.8\\ 8.1
        \end{tabular} & 1000$^{\text{a}}$ & \begin{tabular}{r} 0.13\\ 0.43
        \end{tabular}\\
        
        \bottomrule
    \end{tabular}
    \hspace{3cm}
    \footnotetext{Expected value being extremely conservative. We used references~\cite{doi:10.1063/5.0020019,Jiang:19} for the estimative.}
\end{sidewaystable}

\end{document}